%% file: main.tex
\documentclass[sigconf, anonymous=false]{acmart}

\usepackage{xspace}

\usepackage[linesnumbered,ruled,vlined]{algorithm2e}
\usepackage{bm}
\usepackage{amsmath,amsfonts,amsthm}
\usepackage{mathtools}
\usepackage{makecell}
\usepackage{bm}
\usepackage{subcaption}
\usepackage{multirow}
\AtBeginDocument{%
  }

\setcopyright{none}
\renewcommand\footnotetextcopyrightpermission[1]{}

\newcommand{\ours}{HP-MDR\xspace}

\begin{document}

\title{\ours: High-performance and Portable Data Refactoring and Progressive Retrieval with Advanced GPUs}

\author{Yanliang Li}
\email{leonli@uoregon.edu}
\affiliation{%
  \institution{University of Oregon}
  \state{OR}
  \country{USA}
}

\author{Wenbo Li}
\email{Wenbo.Li@uky.edu}
\affiliation{%
  \institution{University of Kentucky}
  \state{KY}
  \country{USA}
}

\author{Qian Gong}
\email{gongq@ornl.gov}
\affiliation{%
  \institution{Oak Ridge National Laboratory}
  \state{TN}
  \country{USA}
}

\author{Qing Liu}
\email{qing.liu@njit.edu}
\affiliation{%
  \institution{New Jersey Institute of Technology}
  \state{NJ}
  \country{USA}
}

\author{Norbert Podhorszki}
\email{pnorbert@ornl.gov}
\affiliation{%
  \institution{Oak Ridge National Laboratory}
  \state{TN}
  \country{USA}
}

\author{Scott Klasky}
\email{klasky@ornl.gov}
\affiliation{%
  \institution{Oak Ridge National Laboratory}
  \state{TN}
  \country{USA}
}

\author{Xin Liang}
\email{xliang@uky.edu}
\affiliation{%
  \institution{University of Kentucky}
  \state{KY}
  \country{USA}
}

\author{Jieyang Chen}
\email{jieyang@uoregon.edu}
\affiliation{%
  \institution{University of Oregon}
  \state{OR}
  \country{USA}
}

\renewcommand{\shortauthors}{Trovato et al.}

\begin{abstract}
Scientific applications produce vast amounts of data, posing grand challenges in the underlying data management and analytic tasks. 
Progressive compression is a promising way to address this problem, as it allows for on-demand data retrieval with significantly reduced data movement cost. 
However, most existing progressive methods are designed for CPUs, leaving a gap for them to unleash the power of today's heterogeneous computing systems with GPUs. 
In this work, we propose \ours, a high-performance and portable data refactoring and progressive retrieval framework for GPUs. 
Our contributions are three-fold: (1) We carefully optimize the bitplane encoding and lossless encoding, two key stages in progressive methods, to achieve high performance on GPUs; (2) We propose pipeline optimization and incorporate it with data refactoring and progressive retrieval workflows to further enhance the performance for large data process; (3) We leverage our framework to enable high-performance data retrieval with guaranteed error control for common Quantities of Interest; (4) We evaluate \ours and compare it with state of the arts using five real-world datasets. Experimental results demonstrate that \ours delivers up to $6.6\times$ throughput in data refactoring and progressive retrieval tasks. It also leads to $10.4\times$ throughput for recomposing required data representations under Quantity-of-Interest error control and $4.2\times$ performance for the corresponding end-to-end data retrieval, when compared with state-of-the-art solutions. 
\end{abstract}





\maketitle

\input{tex/introduction}
\input{tex/related}

\input{tex/overview}
\input{tex/bpencoding}

\input{tex/lossless}

\input{tex/qoi}

\input{tex/evaluation}
\input{tex/conclusion}

\bibliographystyle{ACM-Reference-Format}
\bibliography{ref}

\end{document}

%% file: tex/introduction.tex
\section{Introduction}\label{sec:introduction}

With the recent deliveries of exascale computing systems~\cite{aurora, frontier, elcaptain}, scientific applications are producing an unprecedented amount of data that overwhelms the storage and data transfer systems. 
This poses grand challenges in the design and development of exascale data management systems, necessitating the need for efficient and effective data reduction. 

Error-controlled lossy compression is a direct way to address the scientific data challenge, as it can significantly reduce the size of scientific data while enforcing user-specified error controls. 
It has developed rapidly in the last decade, and has been widely deployed in a broad range of application domains, including climatology~\cite{baker2016evaluating}, cosmology~\cite{pulido2019data}, fusion~\cite{cappello2020fulfilling}, and artificial intelligence~\cite{underwood2024understanding}.  
Nonetheless, error-controlled lossy compression has a severe limitation that prevents its broader adoption in science: it provides only a single precision, although the data may be used for different scientific analytics with diverse precision requirements. 
This leads to a dilemma for scientists when choosing the proper error control. 
On the one hand, strict error control may ensure data fidelity for most downstream tasks, but it yields limited benefits in data transfer and storage. 
One the other hand, loose error control can significantly mitigate the pressure on data movement, but it may produce wrong results for downstream tasks that require high precision. 

Data refactoring with error-controlled progressive retrieval~\cite{li2019vapor, liang2021error, magri2023general, wu2024error} has been recently proposed and regarded as an alternative way to manage scientific data. 
Similar to progressive compression with JPEG/JPEG2000~\cite{wallace1992jpeg, rabbani2002jpeg2000} in the image processing community, these approaches refactor data into different precision/resolution segments.
During retrieval, these segments are used to reconstruct the data toward user-specified error control in a progressive and incremental fashion. 
While this does not reduce the data size for storage, it significantly improves the efficiency of data retrieval by providing just enough precision on demand. 
Meanwhile, it eliminates the risk of inaccurate data in classic error-controlled lossy compression, as it can provide near-lossless representations that satisfy the requirement for most downstream tasks. 

Although several progressive methods~\cite{li2019vapor, liang2021error, magri2023general} have been proposed in literature, most of them are designed and optimized for CPU architectures. 
Nonetheless, almost all recent leadership computing facilities are highly heterogeneous with modern GPUs, leaving a significant gap for progressive methods to fully unleash the computational power of these systems.  
This situation is further exacerbated by the diverse GPU architectures on those systems (e.g., NVIDIA GPUs on Summit~\cite{summit}, AMD GPUs on Frontier~\cite{frontier}, and Intel GPUs on Aurora~\cite{aurora}), as a tailored implementation on one system may not work for the others. 
Because of their fundamental differences in architecture, a data refactoring pipeline may use a different algorithm variant on different processor types for optimized performance.
Such differences can cause data portability challenges: data refactored by one type of processor cannot be reconstructed by another type of processor with a guarantee.
Due to the portability challenges, users are forced to use the most compatible processor to ensure data are still retrievable on future systems that use different architectures.  
However, the most compatible processors, such as a single-core CPU, cannot guarantee the best performance.

In this work, we propose a high-performance and portable framework -- \ours, implementing progressive methods on exascale systems with advanced GPUs. 
In particular, we propose end-to-end portable refactoring and reconstruction pipelines with highly optimized register block-based bitplane encoding and hybrid lossless encoding strategies.
To this end, we construct a pipeline to compose PMGARD~\cite{liang2021error}, a state-of-the-art progress method, and further design an execution workflow to enable progressive retrieval with guaranteed error control on Quantities of Interest (QoIs), which represents derived information of most interest to application scientists. 
In summary, our contributions are as follows.

\begin{itemize}
    \item Based on the algorithmic characteristics of bitplane encoding and many-core architectures, we thoroughly study several optimized parallel bitplane encoding strategies, and we design a highly optimized bitplane encoder that accelerates encoding by $2.1\times$ and decoding by $8.3\times$ on modern GPUs while providing portability across architecture types.
    
    \item To adapt to the diverse compressibility of bitplanes, we build a hybrid lossless compression that accelerates bitplane compression throughput by $3.1\times$ with only $8\%$ data retrieval overhead on average.

    \item We build end-to-end data refactoring and reconstruction pipeline on GPU. To further improve GPU utilization, we propose a highly optimized pipeline optimization that accelerates end-to-end refactoring by $1.43\times$ and reconstruction by $1.83\times$ on average. We further incorporate them with multilevel data decomposition algorithms for high-performance data refactoring and retrieval on GPUs, yielding over $6.6\times$ throughput over existing approaches. 
    
    \item We leverage our optimized encoding algorithm and pipeline to enable progressive retrieval with guaranteed error control under common Quantities of Interest (QoIs), leading to $10.4\times$ throughput for data recomposition and $4.2\times$ performance for the underlying end-to-end data retrieval.  

\end{itemize}

The rest of the paper is organized as follows. In Section~\ref{sec:related}, we discuss the related work on scientific data compression and progressive methods. In Section~\ref{sec:overview}, we provide an overview of the proposed framework. 
We then detail our optimization on bitplane encoding and lossless compression in Section~\ref{sec:bp} and Section~\ref{sec:lossless}, respectively. 
In Section~\ref{sec:qoi}, we describe how to compose PMGARD in our framework and enable error control on downstream QoIs with high performance. 
In Section~\ref{sec:evaluation}, we present our evaluation results with real-world datasets. 
In Section~\ref{sec:conclusion}, we conclude the paper with a vision for future works. 

%% file: tex/related.tex
\section{Related Works}\label{sec:related}
In this section, we review the related works on scientific data compression, which is broadly categorized into error-controlled lossy compression and progressive compression. 

\subsection{Error-controlled lossy compression}
The ever-increasing amount of scientific data imposes grand challenges on the underlying data management and analytic tasks, which cannot be addressed by generic lossless compressors such as GZIP~\cite{gzip} and ZSTD~\cite{zstd} due to their limited compression ratios. 
Meanwhile, error-controlled lossy compression~\cite{lindstrom2006fast, lakshminarasimhan2013isabela, lindstrom2014fixed, sz17, liang2018error, zhao2021optimizing, ainsworth2019multilevel, ainsworth2020multilevel, liang2021mgard+, liang2022sz3} has been evolving as a promising solution because it significantly reduces data size while enforcing user-specified error controls that are essential for scientific applications. 

Error-controlled lossy compressors can be broadly categorized as prediction-based and transform-based ones. 
SZ~\cite{sz17, liang2018error, zhao2021optimizing, liang2022sz3} is one of the most widely used prediction-based lossy compressors. 
It relies on various predictors, including Lorenzo~\cite{ibarria2003out} and splines~\cite{zhao2021optimizing}, to decorrelate the data, followed by a linear-quantization stage to reduce the entropy while ensuring error control. The quantized data is then fed to lossless encoders such as Huffman~\cite{huffman1952method} and ZSTD for further size reduction. 
ZFP~\cite{lindstrom2014fixed} is a typical transform-based lossy compressor that leverages block transform for decorrelation. 
In particular, it divides the original data into independent blocks, and performs a near-orthogonal transform after fixed-point alignment in each block. 
The transform coefficients are then encoded with an efficient embedded encoding algorithm and concatenated for storage. 
MGARD~\cite{ainsworth2018multilevel, ainsworth2019multilevel, ainsworth2020multilevel, liang2021mgard+} is another popular lossy compressor lying in the middle, which features rigorous error control theories on raw data and downstream Quantities of Interest (QoIs). 
It establishes a novel decomposition algorithm based on finite element analysis and wavelet theories, followed by linear quantization and lossless encoding stages similar to those of SZ. 
In addition to these mainstream compressors, scientific data compression has also been advanced by many other methods, such as wavelet transforms (SPERR~\cite{sperr}), singular value decomposition (TTHRESH~\cite{ballester2019tthresh}), and deep learning (AE-SZ~\cite{liu2021exploring}). 

There has been a growing trend in implementing and optimizing error-controlled lossy compressors on GPUs to facilitate their use on heterogeneous leadership computing facilities. 
Nonetheless, adaptions of compression algorithms are usually required to better unleash the parallel processing power of advanced GPUs due to either inherently sequential operations or underoptimized design. 
For instance, cuSZ~\cite{tian2020cusz} adopts a dual-quantization design to eliminate the dependency in Lorenzo prediction, achieving decent compression ratios with high throughput. 
GPU-MGARD~\cite{chen2021accelerating} optimizes three critical kernels for efficient grid, linear, and iterative processing, leading to significant speedup over a naive porting version. 
Recently, the throughput of the scientific data compression kernel has been pushed to hundreds of GB/s on NVIDIA GPUs~\cite{huang2024cuszp2}. 

One critical problem of error-controlled lossy compressors, along with their GPU variations, is that they only provide a single error bound and assume this accuracy could be sufficient for all subsequent data analytics. 
However, this could hardly be true due to the diverse requirements in scientific analytics, and scientists have to choose a conservative error bound during compression to ensure sufficient accuracy in the data. 
This usually leads to limited compression ratios, which limits the use of error-controlled lossy compressors in practice.

\subsection{Progressive compression}
Unlike error-controlled lossy compression, progressive compressors~\cite{li2019vapor, liang2021error, magri2023general, wu2024error} store the data in a near-lossless fashion and provide on-demand access during retrieval. 
Although this may not improve the writing performance, it significantly reduces the data movement time during retrieval. 
This aligns well with the characteristics of scientific data, which is usually written once and retrieved multiple times for diverse analytics. 

Progressive compression was first adopted in the image processing community, where JPEG/JPEG2000~\cite{wallace1992jpeg, rabbani2002jpeg2000} divides the image into multiple scans to provide different quality levels. This allows for progressive rendering that starts with low quality but gradually refines with additional data, leading to a better experience for displaying images in webpages.  
PMGARD~\cite{liang2021error} borrowed this concept and took it to the scientific data domain by coupling MGARD decomposition theories and bitplane encoding algorithms to provide near-lossless data refactoring and error-controlled retrieval. 
It was recently improved to provide error control on a set of derived quantities of interest (QoIs)~\cite{wu2024error}, significantly expanding its use in practice. 
Another family of progressive methods relies on existing error-control lossy compressors to achieve progressiveness with error control~\cite{magri2023general}. 
In particular, they iteratively compress the original data and the corresponding residues with off-the-shelf compressors using a set of progressively decayed error bounds. 

Despite the promising usage of progressive compression in scientific data management, little effort has been made to implement the entire procedure on advanced GPUs.
While the iterative procedure~\cite{magri2023general} could be easily ported to GPUs using GPU-based error-controlled lossy compressors, it suffers from low efficiency because GPU-based lossy compressors are not adept at dealing with residue compression, especially when the error bound is relatively low. 
This leaves a significant gap for deploying progressive compression methods in the leadership computing facilities. 

In this work, we propose and develop \ours, a high-performance portable data refactoring and progressive retrieval framework for advanced GPUs. 
In particular, we propose a set of tailored optimizations to significantly improve the performance of bitplane encoding and lossless compression of the encoded bitplanes, which are identified as the primary performance bottlenecks. 
Based on existing bitplane encoding works~\cite{li2019bstc, lindstrom2025zfp}, we thoroughly optimized parallel bitplane encoding on GPUs to double the performance compared with the best existing works.
To this end, we couple our methods with GPU-MGARD to form an end-to-end data refactoring and progressive retrieval pipeline, and we further enable error controls on downstream QoIs for practical usage.

%% file: tex/overview.tex
\section{System Overview}\label{sec:overview}

\begin{figure}[t]
\centering
\vspace*{-1em}
\includegraphics[width=\linewidth]{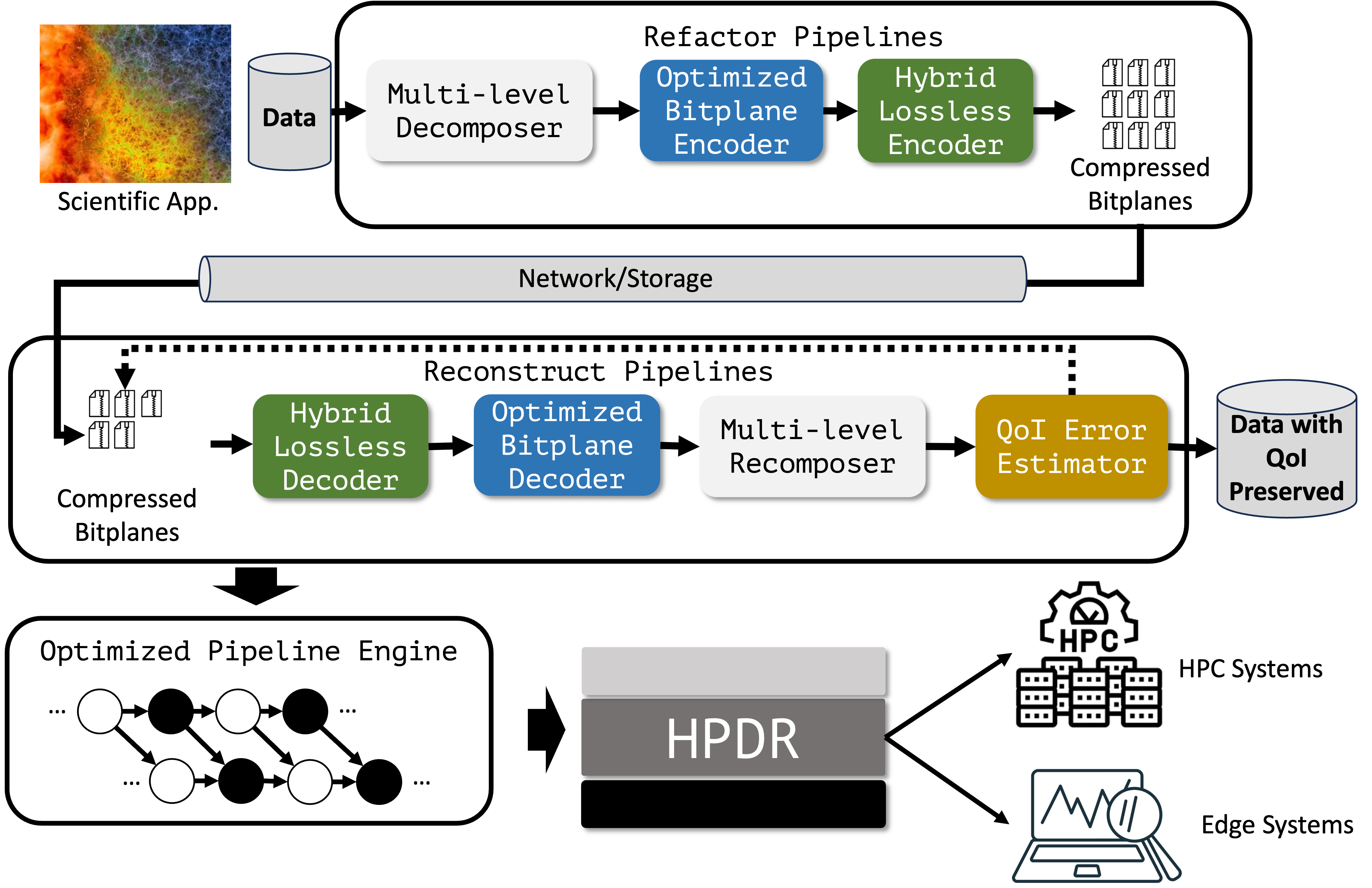}
\caption{High-perf. portable multi-precision data refactoring framework (\ours)}
\label{framework}
\vspace*{-1em}
\end{figure}

Figure~\ref{framework} shows the system overview of the \ours framework.
The input data is first decomposed using the multi-level decomposer, then encoded into bitplanes, and finally compressed into reduced form. 
Part of the compressed bitplanes can be used to reconstruct data with error control via the reverse operations.
In this pipeline, only the multi-level (re)decomposer is accelerated using GPU~\cite{chen2021accelerating}, while the rest are done on CPU, which forms performance bottlenecks. 
In this work, we aim to design the first end-to-end GPU accelerated refactoring and reconstruction pipelines. We first study the main challenges of accelerating and parallelizing the bitplane encoding and propose optimized encoding kernels that maximize the GPU utilization for various encoding workloads.
Then, we design a hybrid lossless compressor that adaptively leverages Huffman and Run-Length Encoding (RLE) to compress each bitplane efficiently.
To further optimize end-to-end performance, we design pipeline optimizations that overlap CPU-GPU memory copy with computation for both refactoring and reconstruction pipelines.
Furthermore, we leverage the pipeline optimization to build the first high-performance progressive retrieval pipeline with QoI error control on GPUs. 
Finally, the end-to-end computations are implemented on top of the High Performance Portable Data Reduction framework~\cite{chen2025hpdr}, which enables portability across various types of GPU and CPU architectures.

%% file: tex/bpencoding.tex
\section{Bitplane Encoding}\label{sec:bp}
Bitplane encoding is a core component in many progressive compression frameworks, as it can provide very fine-grained precision decomposition to enable progressiveness.
The encoding process is illustrated in Algorithm~\ref{alg:bitplane_encoding}. Given a decomposed input array $Q$, the algorithm first aligns all elements by exponent to ensure consistent bitplane boundaries.
This alignment is performed with respect to the global maximum exponent across all elements, so that the most significant bits (MSBs) are preserved throughout the batch.
It then iterates through $B$ bitplanes from the most to the least significant bits. 
For each bitplane, it extracts the corresponding bit from every element, stores the results as 1 bit of the encoded bitplane and finally writes the entire encoded bitplane to the output stream.

\begin{algorithm}[t]
\caption{Bitplane Encoding Overview}
\label{alg:bitplane_encoding}
\KwIn{Decomposed data array $Q$ of length $N$, target bitplane count $B$}
\KwOut{Bitplane-encoded stream $S$}

\tcc{1. Shift all values to align MSBs}
$aligned\_Q \leftarrow \text{AlignExponent}(Q)$

\tcc{2. Loop over B bitplanes}
\For{$b \leftarrow B-1$ \KwTo $0$} {
    $encoded\_bitplane \leftarrow$ empty array of length $N$ \tcp*[r]{Initialize empty bitplane}
    \ForPar{$i \leftarrow 0$ \KwTo $N - 1$} {
        $bit \leftarrow (aligned\_Q[i] \gg b) \mathbin{\&} 1$ \tcp*[r]{Extract bit-$b$}
        $encoded\_bitplane[i] \leftarrow bit$ \tcp*[r]{Store bit-$b$}
    }
    \texttt{write\_bitplane}$(S, encoded\_bitplane)$ \tcp*[r]{Flush one bitplane}
}
\end{algorithm}

Although it has low algorithmic complexity and arithmetic intensity, designing and optimizing parallel encoding algorithms for many-core architectures such as modern GPUs is non-trivial.
This is because (1) it is hard to choose a proper parallelization strategy that maximizes both GPU occupancy and memory access efficiency, and (2) fine-grain parallelization can incur large inter-thread communication overhead. Previously, several GPU parallel bitwise processing algorithms have been proposed~\cite{li2019bstc, lindstrom2025zfp} that potentially can be used to build bitplane encoding. However, they have not been thoroughly optimized and compared in the context of bitplane encoding. In this work, we explore and compare three optimized bitplane encoding designs.

\begin{figure}[t]
\centering
\includegraphics[width=\columnwidth]{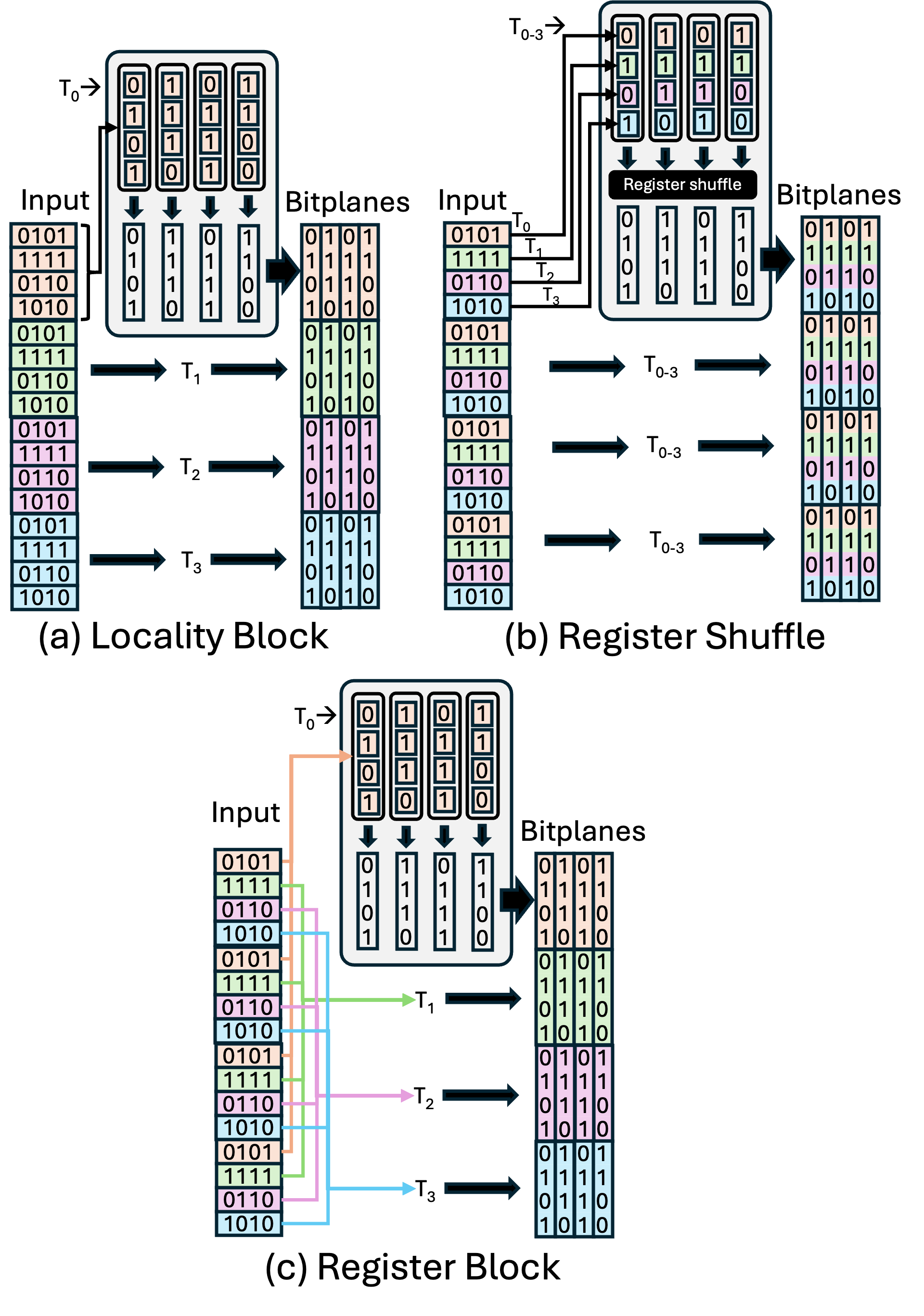}
\vspace*{-2em}
\caption{Three optimized designs for bitplane encoding}
\label{fig:bitplane encoding}
\vspace*{-2em}
\end{figure}

\subsection{Bitplane encoding with locality block}
Our first design takes inspiration from the ZFP lossy compressor~\cite{lindstrom2025zfp}, where bitplane coding is one key step in its compression pipeline.
In this design, a relative coarse parallelization strategy is used where each thread encodes a $4^D$ block consisting of neighboring elements. 
In our design, we group every contiguous $B$ input element into a locality block and encode their bits into the same bitplane data block.
Similar to ZFP, each thread is assigned to encode one locality block when parallelized on GPUs.
Figure~\ref{fig:bitplane encoding} (a) illustrates a toy example of encoding 4 bitplanes with each locality block containing 4 elements.
The figure shows four threads (i.e., $T_0 ... T_3$), with each thread encoding 4 input data and storing the encoded bitplanes independently.
This design's main advantage is that it preserves the locality of the input elements in the encoded bitplanes, which can help preserve the bitplane's compressibility.
For example, neighboring coefficients tend to have a closer value range, and their higher bits tend to be similar, resulting in more contiguous 0 or 1 bits in the encoded bitplanes.
This design also provides a fair amount of parallelism when the input size is large, and it does not involve any inter-thread communications.
Moreover, the data access pattern for storing encoded bitplanes can be fully coalesced.
The main drawback of this design is that the memory load pattern is not coalesced.
For smaller block sizes, this issue can be mitigated through the L2 cache.
However, smaller blocks tend to reduce the amount of work per thread, which impacts inter-instruction parallelism.
So, finding the suitable blocks is the key optimization strategy for this design.

\subsection{Bitplane encoding with register shuffling}
\begin{figure}[t]
\centering
\vspace*{-1em}
\includegraphics[width=\linewidth]{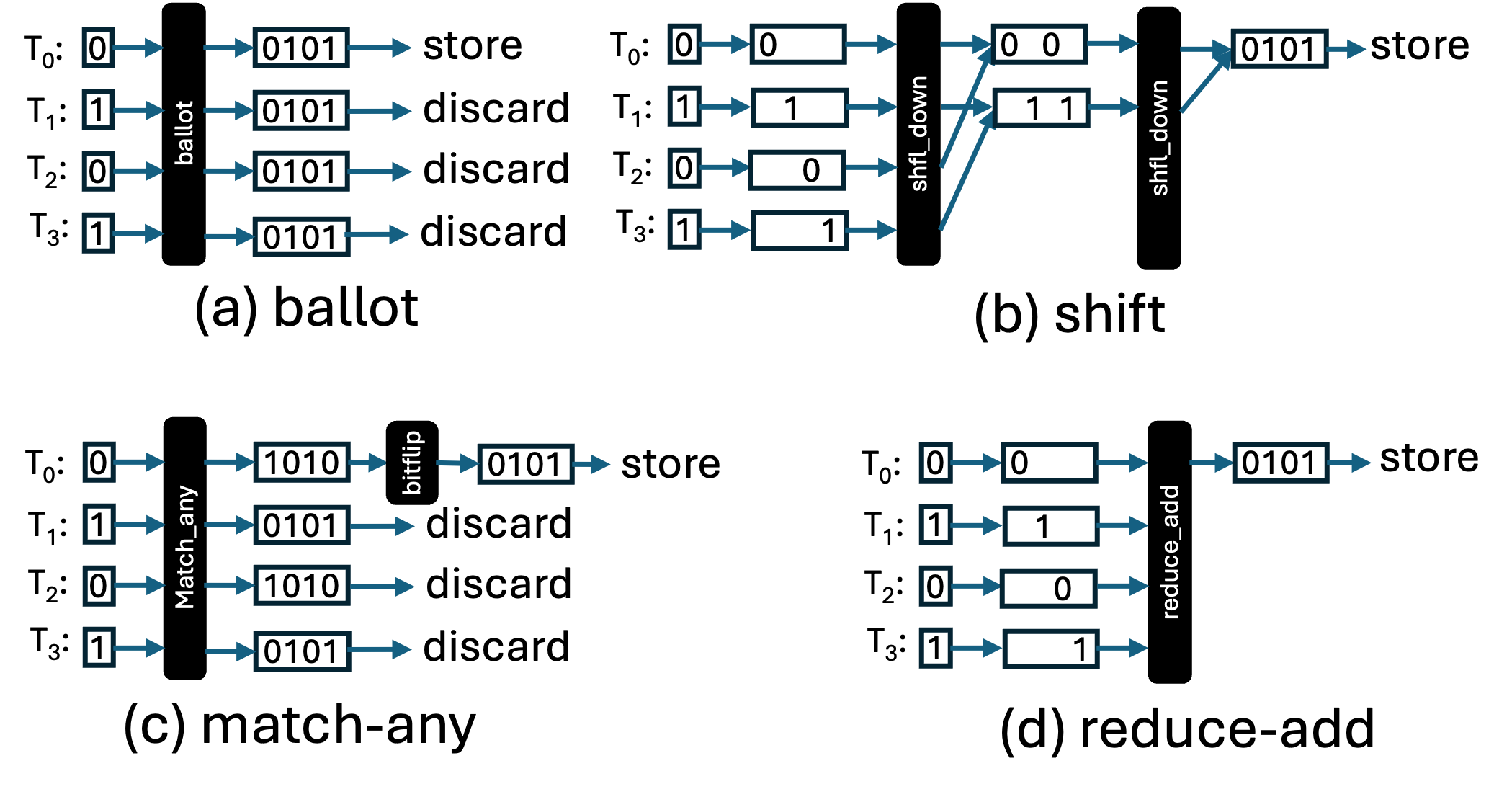}
\vspace*{-2em}
\caption{Four encoding designs using register shuffling}
\label{fig:shuffle-encoding}
\vspace*{-1em}
\end{figure}

For smaller input sizes, the locality block design suffers from low parallelism (e.g., parallelism = n/B, where n is the input size).
An alternative approach to improving concurrency is having each thread load one element from the input.
However, this creates a problem for encoding as threads loading neighboring elements must exchange bits to encode bitplanes.
In this design, we extend the register shuffling-based bit-matrix transpose algorithm proposed in \cite{li2019bstc} to enable bit exchange.
After each thread loads its element, each bit is extracted and shared with others via GPU register shuffling.
Our design differs from \cite{li2019bstc} in the following two ways: (1) we need to consider memory load and store pattern in addition to bit exchange; (2) \cite{li2019bstc} only uses the warp ballot operation. We explore the design with four different register-shuffling instructions, as shown in Figure~\ref{fig:shuffle-encoding}.

In the ballot approach, each thread sends its bit as a predicate value, and although not needed, all threads get the voting results in the end. In this case, only the thread responsible for storing this particular bitplane keeps the results, and the rest of the threads discard the results.
The ballot approach uses the fewest instructions, but it does introduce unnecessary communication as results are broadcast to all threads.
The shift approach leverages a classic tree reduce, and only the storing thread obtains the final encoded bitplane. Although it incurs less communication, it requires multiple rounds of shift operations.
Match-any behaves similarly to ballot. The only difference is that the result may need an additional bit flip operation if the storing thread holds bit 0.
Reduce-add behaves similarly to the shift approach, except the reduction process is repeated with the reduce instruction, which may exploit dedicated hardware optimizations.

\subsection{Bitplane encoding with GPU register block}
Both the locality block and register shuffling approach contain certain performance degradation factors; The locality block brings non-fully coalesced memory access pattern, and the register shuffling approach requires extensive inter-thread communication.
Our third design explores a fully memory coalesced and communication-free bitplane encoding approach.
As shown in Figure~\ref{fig:bitplane encoding} (c), we propose a register block-based approach.
Specifically, to avoid any inter-thread communication, we let each thread encode $B$ elements.
To achieve fully coalesced memory access, instead of loading contiguous elements, we let each thread load interleaved elements, such as neighboring threads always loading consecutive elements to achieve a coalesced pattern.
After loading all data, each thread caches the intermediate data in its own register block and performs encoding independently.
Moreover, as all data are in the register blocks, GPUs can fully exploit instruction-level parallelism for better overall throughput.
The main drawback of this approach is that bit correlation is not preserved through the encoding process as data are used in an interleaved manner. 
However, such impact is only limited to each $warp\_size \times B$ region where an interleaved access pattern occurs.
So, the compressibility degradation is limited.

%% file: tex/lossless.tex
\section{Lossless Encoding}\label{sec:lossless}
Lossless encoding is applied to the encoded bitplanes for further size reduction without information loss. 
It is a crucial step in data refactoring and progressive retrieval framework as its efficiency directly influences the data size and the underlying data movement cost. 
In \ours, we consider the adaptive use of three core lossless methods ---Huffman coding, Run-Length Encoding, and Direct Copy--- to achieve the best efficiency. 
In the following texts, we first introduce the three lossless encoding techniques, followed by our hybrid lossless compression algorithm. 


\subsection{Lossless encoding technique}

\textbf{Huffman encoding (Huffman)}~\cite{tian2021revisiting} is an entropy-based method that assigns shorter binary codes to more frequent symbol. In our framework, it is particularly effective for higher-order bitplanes, where the frequent distribution of symbols is heavily concentrated on a few values, especailly zeros. We adopt a parallel, GPU-optimized implementation to efficiently compress large-scale bitplane blocks.

\textbf{Run-Length Encoding (RLE)}~\cite{balevic2009fine} compresses sequences of repeated values by encoding them as (value, count) pairs. This method performs well on lower-order bitplanes, where long runs of zeros frequently occur due to quantization and truncation. Compared to Huffman, RLE achieves lower computational overhead and excels in capturing structured sparsity.

\textbf{Direct Copy (DC)} bypasses compression entirely and stores the bitplane data as-is. This strategy is applied when the data size is small or the bitplane is not sufficiently compressible, thus avoiding unnecessary encoding overhead while maintaining high throughput during progressive retrieval.

Huffman and RLE are the primary choices due to their complementary strengths in exploiting different types of redundancy, while DC serves as a lightweight fallback when neither method is effective.
These three techniques collectively form the foundation of our hybrid lossless encoding strategy.

\subsection{Hybrid lossless compression}

To further optimize storage and retrieval performance, we propose an \textit{Hybrid Lossless Compression} that dynamically selects the most appropriate method for each group of bitplanes. 

Every four consecutive bitplanes (a configurable unit) are merged and evaluated for compressibility. We estimate the potential compression ratio of both Huffman and RLE using light-weight predictors, and then choose the most suitable encoding method according to size and compression ratio thresholds. The full decision logic is presented in Algorithm~\ref{alg:hybrid_lossless_compression}.

\begin{algorithm}[t]
\caption{Hybrid Lossless Compression Strategy}
\label{alg:hybrid_lossless_compression}
\KwIn{Bitplane array \texttt{B}, group size \texttt{m}, size threshold \texttt{$T_s$}, CR threshold \texttt{$T_{cr}$}}
\KwOut{Compressed bitplane array \texttt{C}}

\texttt{N} $\gets$ total number of bitplanes in \texttt{B} \\
\For(\tcp*[f]{loop over bitplane groups}){\texttt{i} $\gets$ 0 $\to$ \texttt{N} $-1$}{
  \texttt{G} $\gets$ merge \texttt{B[i..i+m-1]} \tcp*{merge $m$ bitplanes}  
  \texttt{S} $\gets$ size of \texttt{G} \\
  \If{\texttt{S} $>$ \texttt{$T_s$}}{
    \texttt{r\_H} $\gets$ estimate CR by Huffman on \texttt{G} \\
    \texttt{r\_R} $\gets$ estimate CR by RLE on \texttt{G} \\
    \eIf{\texttt{r\_H} $>$ \texttt{$T_{cr}$}}{
      \texttt{C[i]} $\gets$ \texttt{HuffmanEncode(G)} \tcp*{Use Huffman}
    }{
      \eIf{\texttt{r\_R} $>$ \texttt{$T_{cr}$}}{
        \texttt{C[i]} $\gets$ \texttt{RLEEncode(G)} \tcp*{Use RLE}
      }{
        \texttt{C[i]} $\gets$ \texttt{DirectCopy(G)} \tcp*{Use DC}
      }
    }
  }
  \Else{
    \texttt{C[i]} $\gets$ \texttt{DirectCopy(G)} \tcp*{Use DC}
  }
  \For{\texttt{j} $\gets$ 1 \KwTo \texttt{m} $-1$}{
    \texttt{C[i + j]} $\gets$ \texttt{EmptyPlaceholder()}
  }
}
\end{algorithm}

The key to the efficiency of our hybrid lossless compression is the accurate yet inexpensive estimation of the compression ratio (CR) for both Huffman encoding and RLE. In the following paragraphs, we delve into the details of each CR estimation method.

\paragraph{Huffman Compression Ratio Estimation}
For Huffman encoding, we first compute a frequency histogram of the symbols in the merged bitplane. Based on this histogram, an optimal Huffman tree is generated, assigning shorter code lengths to more frequent symbols. The estimated bit cost is then calculated as the sum of the products of each symbol's frequency and its corresponding code length. The CR is determined by comparing the original data size to this estimated cost (adjusted for any constant overhead).

\paragraph{RLE Compression Ratio Estimation}
For RLE, the estimation is based on an efficient scan of the data to mark the beginnings of runs of consecutive identical symbols. We then compute the total run length by summing these markers. The encoding cost for each run is approximated by considering both the fixed cost to store the symbol and the variable cost to encode the run length. The CR is derived by taking the ratio of the original data size to the estimated total bit cost for encoding all the runs.

If either estimation exceeds the thresholds $T_{cr}$, the corresponding encoder is selected. Otherwise, Direct Copy is applied. This logic ensures that encoding effort is only applied when beneficial.

The CR estimation is performed ahead of actual encoding and incurs minimal overhead. Furthermore, to preserve stream alignment and decoding compatibility, placeholder slots are reserved for non-leading bitplanes in each group.

%% file: tex/qoi.tex
\section{Pipeline Optimization and Implementation}\label{sec:qoi}
In this section, we introduce our pipeline optimization, which significantly improves GPU utilization to achieve high end-to-end refactoring and reconstruction performance for large-scale data. 
We then leverage it to construct the first end-to-end data refactoring and progressive retrieval framework on GPUs and enable guaranteed error control for derivable Quantities of Interests (QoIs).

\begin{figure*}[ht]
\centering
\includegraphics[width=0.8\linewidth]{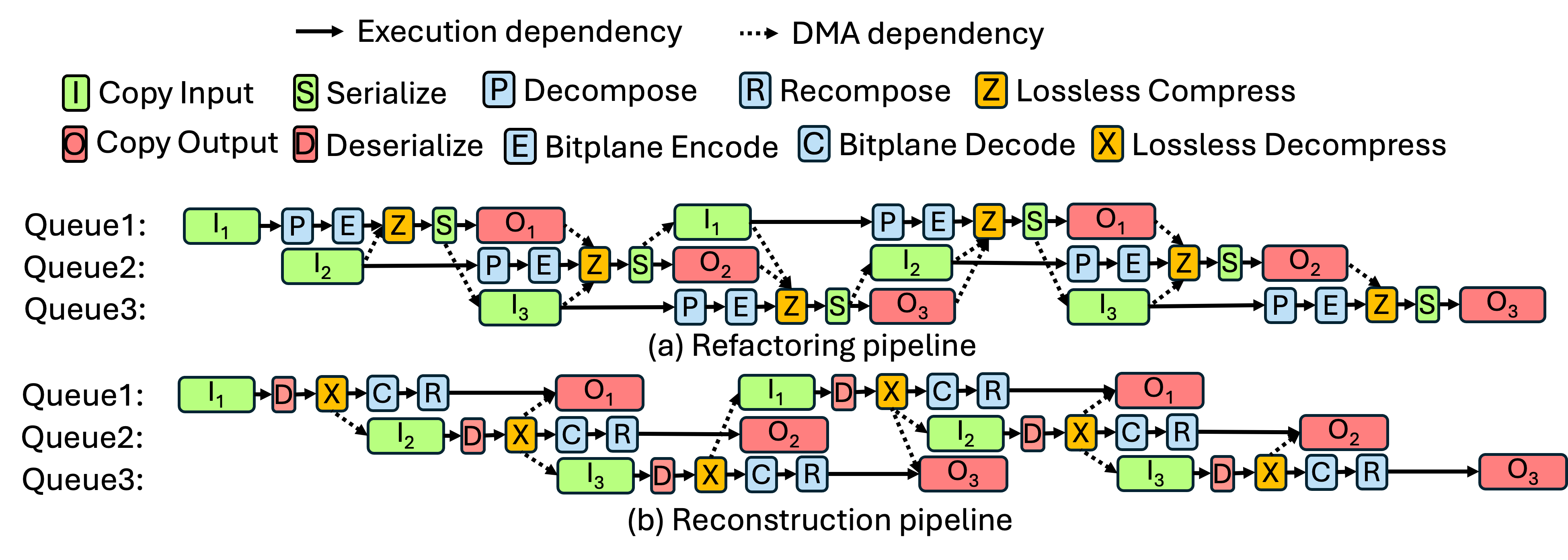}
\vspace*{-1em}
\caption{Optimized refactoring and reconstruction pipeline represented as DAGs that enable data transfer latency hiding. Green: host-to-device copy; Red: device-to-host copy; Blue: compute; Yellow: copy-compute mixed}
\label{pipeline-opt}
\vspace*{-1em}
\end{figure*}
\subsection{Pipeline optimization}
When refactoring or reconstructing a large-scale dataset, the entire data may not be able to fit entirely on GPU memory. In this case, data needs to be decomposed into sub-domains and to be processed sequentially.
Also, when processing multiple variables, each variable also needs to be refactored and reconstructed sequentially.
As pointed out in~\cite{chen2025hpdr}, frequent data copy in and out of GPU devices can incur large overhead for data reduction pipelines.
In this work, we extend the pipeline optimization in~\cite{chen2025hpdr} to refactoring and reconstruction pipeline.

To make the pipeline optimization portable across GPU architectures, we use the Host-Device Execution Model (HDEM)~\cite{chen2025hpdr}, to aid design. In this machine model, one GPU device is equipped with two Direct Memory Access (DMA) engines, each of which can work independently for asynchronous memory copy. They are used to copy data between the application buffer, I/O buffer, and refactoring/reconstruction buffer. The device also has a compute engine to support the concurrent execution of refactoring/reconstruction kernels during data copy operations. 

\subsubsection{Data Refactoring}

Figure~\ref{pipeline-opt} (a) shows our optimized refactoring pipeline.
The refactoring process is pipelined among three queues (1-3). Green boxes represent CPU-to-GPU copy tasks. Red boxes represent GPU-to-CPU copy tasks. Blue boxes are pure computing tasks. Yellow represents mixed memory copy and computing tasks. According to our restrictions, no two tasks with the same color can be executed simultaneously, and a yellow task cannot concurrently execute with any other tasks. Three input/output buffers are: $I_1$/$O_1$, $I_2$/$O_2$, and $I_3$/$O_3$. We assume serialization and deserialization are needed for embedding and extracting metadata after and before computation, which also relies on memory copies. Also, lossless compression and decompression contains computation and data copies between CPU and GPU due to its internal serialization and deserialization process, which are color-coded in yellow.
To ensure refactoring correctness, we enforce execution ordering with solid arrows.
Additionally, to hide the memory copy latency, we prefetch the next input while refactoring the current sub-domain.
To avoid delaying the current execution, prefetch needs to be done during multi-level decompositions and bitplane encoding and finished before lossless compression, so we enforce additional dependencies between $ I\rightarrow Z$.
Also, the prefetching should be done as soon as its DMA becomes available (after serilization), so we add another dependency between $S\rightarrow I$.
Finally, to hide the latency of copying refactoring data back to CPU, we let it overlap with multi-level decompositions and bitplane encoding and prefetch.



\subsubsection{Progressive Retrieval}
Figure~\ref{pipeline-opt} (b) shows our optimized reconstruction pipeline.
Similar to the refactoring pipeline, we add additional dependencies to maximize the latency hiding while minimizing potential delay to the original reconstruction pipeline.
To perfect refactored input data, we delay its initiation until we are done with deserialization and lossless decompression ($ X \rightarrow I$). This is because having an early data prefetch can delay the current pipeline due to the conflict used with CPU-to-GPU DMA.
Also, similarly, the GPU-to-CPU memory copy for storing reconstructed data of the last iteration can also delay the current process, so we delay it until we are about to start bitplane decoding and multi-level recomposition ($X \rightarrow O$).



\subsection{Progressive retrieval with QoI error control}
\label{pipeline-qoi}
We further leverage pipeline optimization to efficiently enable multivariate QoI error control during progressive retrieval in \ours based on prior work~\cite{wu2024error}. 
The algorithm is presented in Algorithm~\ref{alg:retrieval_qoi}, with orange statements representing memory operations and blue statements indicating computing operations. 
In particular, the estimated QoI error $\tau'$ is initialized as infinity (line 1), and it will be updated iteratively until its value is less than the target QoI error tolerance $\tau$ (lines 2-10). 
In each iteration, we first fetch the necessary bitplanes and use them to recompose data to their estimated data error bounds in lines 3-7 (initialized as the relative value of $\tau$ over its maximal value multiplied with the value range of data according to~\cite{wu2024error}). 
We then implement the GPU kernels to estimate the supremum of QoI errors and use it to update $\tau'$ based on~\cite{wu2024error}, which consists of the error estimation for several families of base QoIs and some specific operations (line 8). 
Since the target QoIs are computed point-wise with a constant number of operations, this step could be very fast with GPUs. 
If $\tau'$ is greater than $\tau$, we use dedicated methods (to be detailed below) to estimate the next data error bounds to guide the retrieval procedure based on the information we have (lines 9-10). 
Data transfer and recomposition (lines 3-7) are the most time-consuming parts in a single iteration, and our pipeline optimization could effectively overlap them with the proposed pipeline optimization to achieve high throughput. 

\begin{algorithm}[h]
\caption{Progressive retrieval with QoI error control}
\label{alg:retrieval_qoi}
\KwIn{QoI error tolerance $\tau$, initial data error bounds $\{\epsilon_i\}$, encoded bitplanes for all variables $\{\{S_j\}^{(i)}\}$, QoI $\mathcal{Q}$}
\KwOut{Decompressed data with QoI error less than $\tau$}

\tcp{Colors encode {\color{orange}memory} and {\color{blue}computing} operations}  
\tcp{Initialize current QoI error}
$\tau' \gets \infty$ \\
\While{$ \tau' > \tau$ } {
    \tcp{Fetch the first variable}
    {\color{orange}\texttt{copy\_to\_device}($\{S_j\}^{(0)}$, $\epsilon_0$) }\\
    \For{$i \leftarrow 0$ \KwTo $n_v - 2$} {
        \tcp{Fetch the next variable}
        {\color{orange}\texttt{copy\_to\_device}($\{S_j\}^{(i+1)}$, $\epsilon_{i+1}$)}\\ 
        \tcp{recompose current variable}
        {\color{blue}$v_i \gets$ \texttt{recompose}($v_i, \{S_j\}^{(i)}$)}
    }
    \tcp{recompose the last variables}
    {\color{blue}$v_{n_v-1} \gets$ \texttt{recompose}($v_{n_v-1}, \{S_j\}^{({n_v-1})}$)}\\
    \tcp{Estimate QoI errors}
    {\color{blue}$\tau' \gets$ \texttt{estimate\_QoI\_error}($\{v_j\}$, $\{\epsilon_j\}$, $\mathcal{Q}$)}\\
    \If{$\tau' > \tau$}{
        \tcp{Estimate error bounds for all variables}
        {\color{blue}$\{\epsilon_j\} \gets$\texttt{estimate\_next\_eb}($\{v_j\}$, $\{\epsilon_j\}$, $\tau$, $\tau'$, $\mathcal{Q}$)}
    }
}
\Return $\{v_j\}$
\end{algorithm}

We then introduce the methods for estimating the next data error bounds, which is essential to both efficiency and throughput of progressive retrieval with QoI error control. 
Generally speaking, a small number of retrieved bitplanes represent high efficiency (since less data is retrieved), and fewer iterations indicate high throughput (due to less computation).
We explore three methods to perform the estimation in \ours, as detailed below. 

\paragraph{CPU Porting (CP)} We directly port error estimation method from the CPU implementation in~\cite{wu2024error}. In particular, the algorithm first identifies the data point with the maximal estimated QoI error (on GPU), and then iteratively decays the data error bounds and re-evaluates the QoI error for that single data point until the target QoI error tolerance is met (on CPU after transferring necessary information back). This algorithm usually converges to a set of sufficient data error bounds quickly, but it may suffer from over-preservation because the estimation is not accurate due to the use of stale data. This generally leads to redundancy in data retrieval and, thus, suboptimal efficiency. 

\paragraph{Minimal Augmentation (MA)} To address the over-preservation issue in CP, we propose minimal augmentation to obtain a near-optimal retrieval efficiency by fetching data with fine granularity. 
In particular, we directly fetch one more merged bitplane for each variable and update their corresponding data error bounds accordingly. 
Since this method explores the possible combinations of data error bounds at very fine granularity, it could terminate the procedure promptly when sufficient bitplanes are retrieved, leading to high retrieval efficiency.
Nonetheless, it may cost a number of iterations to complete, which negatively impacts the throughput. 

\paragraph{Minimal Augmentation with Proportional Estimation (MAPE)} We further propose to couple minimal augmentation with proportional estimation to reduce the number of iterations needed. 
Given maximal estimated QoI error $\tau'$ and target QoI error tolerance $\tau$, we first check their proportion $p=\tau'/\tau$ to see if they are close enough. 
If $p$ is larger than a threshold $c$, we assume the same proportional relationship on data error bounds and estimate the next data error bound $\epsilon_{i+1}$ as $\epsilon_i / p$, where $\epsilon_{i}$ is the current data error bound; otherwise, we switch to minimal augmentation as the current data representations are very close to the target ones. 
As such, MAPE reduces the number of iterations for convergence while enjoying the benefits of MA, leading to a good tradeoff between retrieval efficiency and throughput. 
Note that we use proportional estimation instead of CP in MAPE because CP easily leads to over preservation, which requires a relatively large $c$ to make the switch. 
This, in turn, will result in a high number of iterations in some instances. 

\begin{figure}[ht]
\centering
\includegraphics[width=0.9\linewidth]{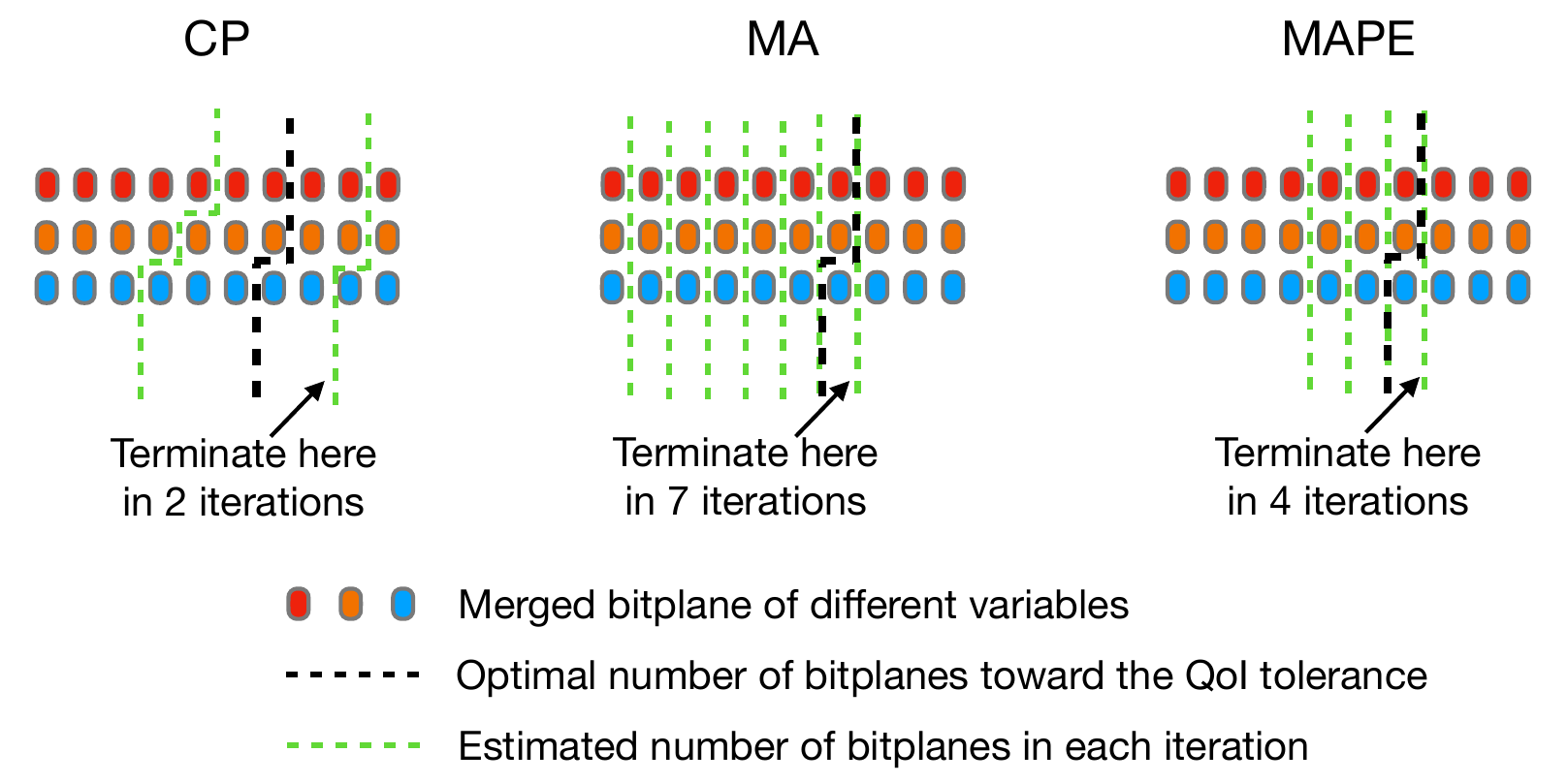}
\caption{Illustration of the three error bound estimation methods: CPU Porting (CP), Minimal Augmentation (MA), and Minimal Augmentation with Proportional Estimation (MAPE). A lower number of bitplanes indicates a smaller retrieval size, and fewer iterations indicate higher throughput.}
\label{fig:eb_est}
\end{figure}

We illustrate the three methods using an example in Fig.~\ref{fig:eb_est}, with a black dashed line indicating the optimal number of bitplanes needed for each variable. 
As shown in the figure, CP can quickly identify the feasible solution, but it may retrieve more bitplanes than needed;
MA yields a near-optimal solution but takes a long time to converge; MAPE has medium efficiency and throughput, which usually provides the best tradeoff. 


%% file: tex/evaluation.tex
\section{Evaluation}\label{sec:evaluation}

\subsection{Experimental setup}
\subsubsection{Experimental environment and datasets}
We conduct experimental evaluations on two systems: Frontier and Talapas.
Frontier is a leadership class exascale supercomputer at Oak Ridge Leadership Computing Facility (OLCF)~\cite{frontier}. It consists of a total of 9,408 computing nodes.
Each compute node has 8 AMD Instinct MI250X GPUs with 64 GB of memory on each GPU and one 64-core AMD EPYC CPU with 512 GB of memory.
Talapas is a heterogeneous cluster system.
Our evaluation is done on one of its GPU computing nodes equipped with 4 NVIDIA H100 GPUs with 80 GB memory on each GPU and two 24-core Intel Xeon CPUs with 1,024 GB memory.
\begin{table}[h]
\caption{Datasets used for evaluation}
\label{dataset}
\resizebox{0.99\columnwidth}{!}{  
\begin{tabular}{|c|c|c|c|c|}
\hline
Dataset     & $n_v$ & Dimensions & Data Type & Size  \\ \hline
NYX & 6 &$512 \times 512 \times 512$        & FP32        & 3 GB \\ \hline
LETKF  & 3 & $98 \times 1200 \times 1200$        & FP32        & 4.9 GB \\ \hline
Miranda  & 3 & $256 \times 384 \times 384$        & FP64        & 1.87 GB \\ \hline
Hurricane ISABEL  & 3 & $100 \times 500 \times 500$        & FP32        & 1.25 GB \\ \hline
JHTDB  & 3 & $1024 \times 2048 \times 2048$        & FP32        & 48 GB \\ \hline
\end{tabular}
}
\vspace*{-1em}
\end{table}





\subsubsection{Baselines} We compare \ours with two baseline works: \textbf{MDR Baseline:} As \ours builds on MDR~\cite{liang2021error}, we include it as a direct baseline. MDR performs multilevel hierarchical decomposition for progressive reconstruction. \textbf{Multi-Component Baselines:} We also evaluate the progressive framework~\cite{magri2023general}, which compresses residual components using different lossy compressors. Selected backends include:  ZFP-GPU~\cite{lindstrom2025zfp} (fixed-rate), MGARD~\cite{chen2021accelerating}, SZ3~\cite{liang2022sz3, zhao2021optimizing, liang2018error}, ZFP-CPU~\cite{lindstrom2014fixed} (fixed-accuracy).


\subsection{Data refactoring and retrieval}

\subsubsection{Bitplane Encoding}

\begin{figure}[t]
\centering
\vspace*{-1em}
\includegraphics[width=0.5\textwidth]{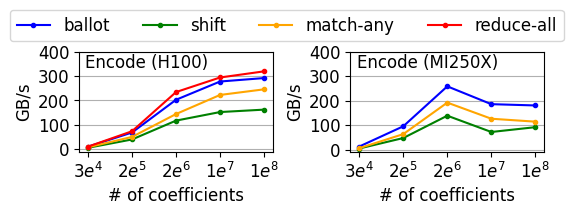}
\vspace*{-3em}
\caption{Bitplane encoding throughput with four types of register shuffle instruction}
\label{exp:bp-shuffle}
\vspace*{-1.5em}
\end{figure}
We first compare bitplane encoding with different register shuffling approaches. We show the performance of encoding 32-bit data into 32 bitplanes and decoding all 32 bitplanes back to 32-bit data with various input sizes.
As shown in Figure~\ref{exp:bp-shuffle}, we evaluate all four register shuffling instructions on H100 and three register shuffling approaches on MI250X since the reduce-all instruction is not implemented on AMD GPUs.
On H100, reduce-all instruction provides the best encoding throughput. Specifically, it improves the encoding performance by up to 15\% compared with start-of-the-art design~\cite{li2019bstc}. This could be due to the existence of dedicated hardware that supports fast reduction.
On MI250X, the ballot outperforms other approaches since it requires the lease amount of instructions.
However, we do observe performance degradation as input size increases that does not exist on H100. This could be due to the architecture difference that causes communication contention to have a more negative impact on AMD GPUs. 

\begin{figure}[ht]
\centering
\vspace*{-1em}
\includegraphics[width=0.5\textwidth]{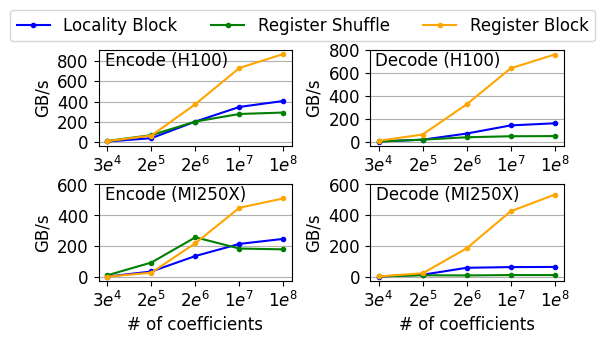}
\vspace*{-3em}
\caption{Bitplane encoding throughput of three parallelization designs}
\label{exp:bp-throughput}
\vspace*{-1.5em}
\end{figure}

Figure~\ref{exp:bp-throughput} shows the throughput comparison of three bitplane encoding parallelization designs on both H100 and MI250X.
We show the performance of encoding/decoding 32-bit data with 32 bitplanes of different input sizes.
For the register shuffling encoder, we use the best-performing instruction throughout the rest of the evaluations.
The evaluation results show that the locality block outperforms register shuffling by $1.4\times$ for encoding and $3.2\times$ for decoding on H100 and $1.4\times$ for encoding and $6.6\times$ for decoding on MI250X.  
Additionally, register block approach outperforms the locality block by $2.1\times$ for encoding and $4.7\times$ for decoding on H100 and $2.1\times$ for encoding and $8.3\times$ for decoding on MI250X.  
The register block provides the highest throughput for both encoding and decoding on both GPUs due to its fully coalesced and communication-free computation for both encoding and decoding. We use this encoding approach for the rest of the evaluations.

\subsubsection{Lossless Encoding}

\begin{figure}[h]
\centering
\vspace*{-1em}
\begin{subfigure}[t]{\columnwidth}
\includegraphics[width=\columnwidth]{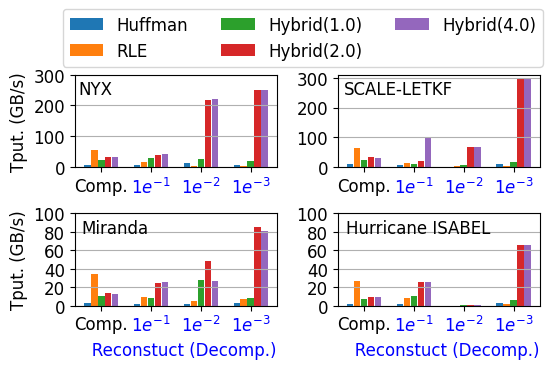}
\vspace*{-1em}
\caption{Throughput}
\end{subfigure}
\vspace*{-1em}
\begin{subfigure}[t]{\columnwidth}
\includegraphics[width=\columnwidth]{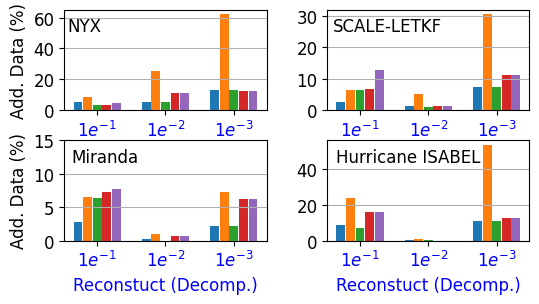}
\vspace*{-1em}
\caption{Incremental data retrieval size}
\end{subfigure}
\vspace*{-0em}
\caption{Comparing performance and compressibility of different lossless compression approaches}
\label{exp:lossless-throughput}
\vspace*{-1em}
\end{figure}

Figure~\ref{exp:lossless-throughput} compares the performance and compressibility of different lossless compression strategies.
Specifically, we compare (1) apply Huffman to all bitplanes (Huffman); (2) apply RLE to all bitplanes (RLE); (3) apply Huffman, RLE, and direct copy hybrid approach with different compression ratio thresholds (Hybrid-rc).
We compare the throughput of the end-to-end lossless (de)compression stage by showing throughput relative to the original data (for compression) and decompressed bitplane size (for decompression).
We also show the incremental data retrieval size when progressively reconstruting to a certain error tolerance. A low retrieval size indicates less I/O cost for progressive data reconstruction.
For the same variable with the same error tolerance, the difference in the retrieval size is only due to the lossless compression's compressibility since the number of bitplanes needed is the same.
As shown in the result, comparing Huffman with others, Huffman brings the least retrieval sizes. However, Huffman has the lower throughput: 5.7 GB/s for compression and 4.8 GB/s for decompression on average. RLE, on the other hand, brings on average 44.4 GB/s for compression and 6.4 GB/s for decompression with 270\% additional data needed for retrieval on average compared with Huffman.  
The hybrid approach brings 15.5 GB/s, 20.8 GB/s, and 22.4 GB/s average compression throughput and 14.1 GB/s, 94.9 GB/s, and 99.8 GB/s average decompression throughput with 8\%, 70\%, and 93\% additional data need compared with Huffman respectively for rc = 1.0, 2.0, and 4.0. 

\subsubsection{End-to-end data refactoring and retrieval}

\begin{figure}[]
\centering
\includegraphics[width=\columnwidth]{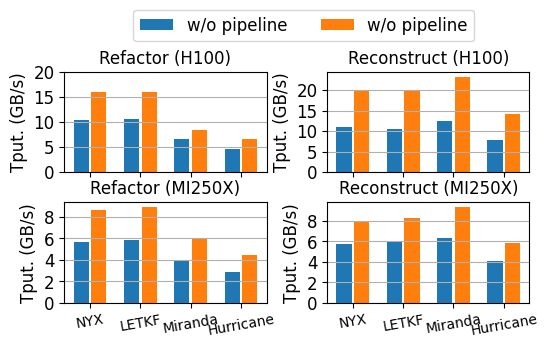}
\vspace*{-2em}
\caption{End-to-End throughput comparison with and without pipeline optimization}
\label{exp:pipeline}
\vspace*{-2em}
\end{figure}

Next, we conduct end-to-end refactoring and reconstruction evaluation.
Figure~\ref{exp:pipeline} compares the end-to-end throughput with and without pipeline optimization.
On H100, the pipeline optimization accelerates refactoring by $1.43\times$ and reconstruction by $1.83\times$ on average.
On MI250X, the pipeline optimization accelerates refactoring by $1.41\times$ and reconstruction by $1.43\times$ on average.

\begin{figure}[t]
\centering
\includegraphics[width=\columnwidth]{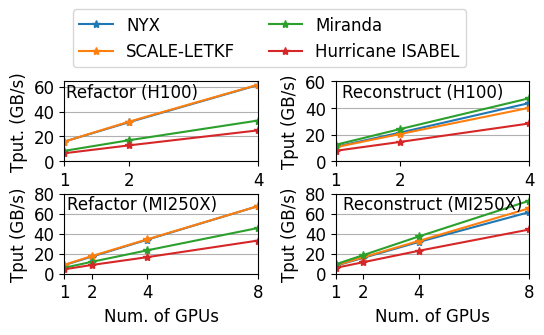}
\vspace*{-2em}
\caption{Scalability on single-node multi-GPU architectures}
\label{exp:scaling}
\end{figure}

Furthermore, we conduct multi-GPU scalability evaluations. We evaluate the end-to-end performance of refactoring and reconstruction in weak-scaling settings.
For H100, we scale up to 4 GPUs. For MI250X, we scale up to 8 GPUs. Figure~\ref{exp:scaling} shows we achieve an average of 95\% and 89\% of the ideal speedup on H100 and MI250X, respectively.

\begin{figure}[t]
\centering
\vspace*{-1em}
\includegraphics[width=\columnwidth]{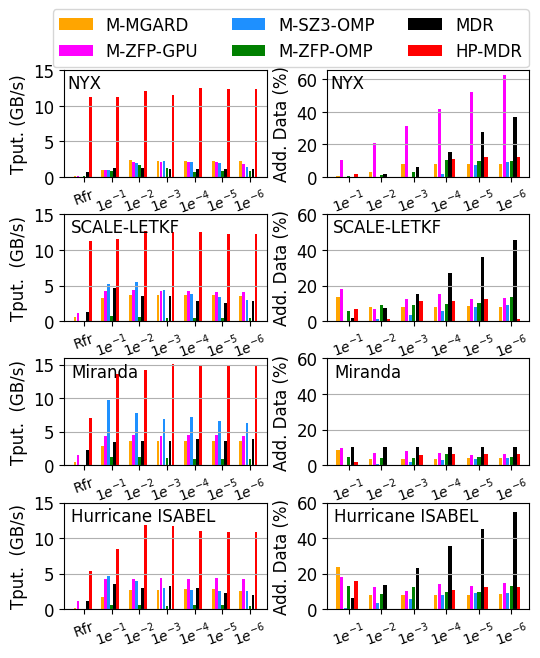}
\vspace*{-2em}
\caption{Comparing \ours with state-of-the-art progressive data retrieve frameworks}
\label{exp:e2e-baseline-compare}
\vspace*{-2em}
\end{figure}

Finally, we compare the end-to-end throughput and retrieval efficiency of our proposed ~\ours with all 5 baselines.
All CPU tests use 32 OpenMP threads; GPU tests run on NVIDIA H100. As shown in Figure~\ref{exp:e2e-baseline-compare}, we consistently outperform all baselines across 4 datasets and a wide range of error tolerances from $10^{-1}$ to $10^{-6}$ on throughput. or example, we achieve an average throughput of 11.9
GB/s while the best baseline (M-MGARD) only obtain
throughput of 1.8 GB/s, meaning that \ours delivers up to $6.61\times$
speedup over it. In the aspect of addition data retrieval, \ours does not yield the smallest retrieval size but remains competitive. For nstance, when reconstructing data on the Miranda dataset, we achieves an average additional retrieval ratio of 4.36\%, which is higher than the best-performing framework (2.19\%) but still better than the overall average across all evaluated baselines (5.55\%). This indicates that while \ours may not minimize retrieval size to the greatest extent, it consistently performs well compared to the majority of existing methods.




\subsection{Progressive retrieval with QoI error control}
All the evaluations in this subsection are performed on the Frontier supercomputer with MI250X GPUs. 
Without loss of generality, we use $V_{total} = V_x^2 + V_y^2 + V_z^2$ as the target QoI. 

\subsubsection{Single-GPU evaluation}
We perform the single-GPU evaluation using velocity fields from NYX (1.5 GB) and mini-JHTDB (6 GB, a cropped region from JHTDB to fit in a single GPU). 
In particular, we compare the efficiency and throughput of all three error bound (EB) estimation methods. 

\begin{figure}[t]
    \centering
    \vspace{-1em}
    \includegraphics[width=\columnwidth]{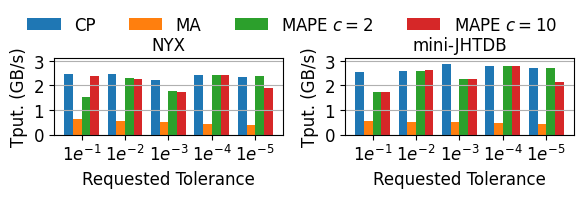}
    \vspace{-2.7em}
    \caption{Overall kernel throughput on the NYX and mini-JHTDB dataset}
    \label{fig:QoI_Single_GPU_TP}
    \vspace{-1em}
\end{figure}

\textit{Retreival efficiency and throughput.} We present the retrieval efficiency on the two datasets in Table~\ref{tab:NYX-perf-qoi} and~\ref{tab:miniJHTDB-perf-qoi}, respectively, and we report the corresponding throughput in Fig.~\ref{fig:QoI_Single_GPU_TP}. 
Overall, it has been observed that Minimal Augmentation (MA) achieves the best bitrates under the majority of requested tolerances with the lowest throughput, while the CPU Porting (CP) achieves the highest throughput under most of the requested tolerances with the worst bitrates. Minimal Augmentation with Proportional Estimation (MAPE) with threshold $c=10$ makes a good tradeoff between ensuring a suboptimal bitrate and maintaining a relatively high throughput, so we use it for the following validation of QoI error control and multi-GPU evaluations. 

\begin{table}[t]
    \centering
    \caption{Bitrate of EB estimation methods on NYX} 
    \label{tab:NYX-perf-qoi}
    \vspace{-1em}
    \resizebox{\columnwidth}{!}{%
    \begin{tabular}{|c|c|c|c|c|c|c|c|c|c|c|c|}
        \hline
        \multirow{1}{*}{\makecell{Method}}
        & \multicolumn{1}{c|}{1E-1} & \multicolumn{1}{c|}{5E-1} & \multicolumn{1}{c|}{1E-2} & \multicolumn{1}{c|}{5E-2} & \multicolumn{1}{c|}{1E-3} & \multicolumn{1}{c|}{5E-3} & \multicolumn{1}{c|}{1E-4} & \multicolumn{1}{c|}{5E-4} & \multicolumn{1}{c|}{1E-5} & \multicolumn{1}{c|}{5E-5} \\
        \cline{2-11}
        \hline
        \texttt{CP} & 6.89 & 6.89 & \textbf{6.89} & 7.49 & 12.57 & 14.90 & 14.90 & 15.20 & 22.90 & 22.90\\
        \hline
        \texttt{MA} & \textbf{4.23} & \textbf{5.99} & 6.90 & \textbf{6.90} & 7.86 & 14.90 & 14.90 & \textbf{14.90} & 22.90 & 22.90\\
        \hline
        \texttt{MAPE}(c=2) & 6.03 & 6.03 & \textbf{6.89} & 7.20 & \textbf{7.82} & \textbf{12.57} & 14.90 & 15.49 & 22.90 & 22.90\\
        \hline
        \texttt{MAPE}(c=10) & \textbf{4.23} & 6.89 & \textbf{6.89} & \textbf{6.90} & \textbf{7.82} & 14.90 & 14.90 & \textbf{14.90} & 22.90 & 22.90\\
        \hline
    \end{tabular}
    }
    \vspace{-0.2em}
    \centering
    \caption{Bitrate of EB estimation methods on mini-JHTDB} 
    \label{tab:miniJHTDB-perf-qoi}
    \vspace{-1em}
    \resizebox{\columnwidth}{!}{%
    \begin{tabular}{|c|c|c|c|c|c|c|c|c|c|c|c|}
        \hline
        \multirow{1}{*}{\makecell{Method}}
        & \multicolumn{1}{c|}{1E-1} & \multicolumn{1}{c|}{5E-1} & \multicolumn{1}{c|}{1E-2} & \multicolumn{1}{c|}{5E-2} & \multicolumn{1}{c|}{1E-3} & \multicolumn{1}{c|}{5E-3} & \multicolumn{1}{c|}{1E-4} & \multicolumn{1}{c|}{5E-4} & \multicolumn{1}{c|}{1E-5} & \multicolumn{1}{c|}{5E-5} \\
        \cline{2-11}
        \hline
        \texttt{CP} & 10.42 & 10.42 & 10.43 & 10.43 & \textbf{11.31} & 18.43 & 18.43 & 18.43 & 26.43 & 26.43\\
        \hline
        \texttt{MA} & \textbf{5.76} & \textbf{5.76} & 10.43 & 10.43 & \textbf{11.31} & 18.43 & 18.43 & 18.43 & 26.43 & 26.43\\
        \hline
        \texttt{MAPE}(c=2) & 6.82 & 10.42 & \textbf{10.42} & 10.43 & 11.38 & 18.76 & 18.43 & 18.43 & 26.43 & 26.43\\
        \hline
        \texttt{MAPE}(c=10) & 6.82 & 8.42 & \textbf{10.42} & 10.43 & 11.38 & \textbf{16.09} & 18.43 & 18.43 & 26.43 & 26.43\\
        \hline
    \end{tabular}
    }
\end{table}

\begin{figure}[t]
    \centering
    \includegraphics[width=\columnwidth]{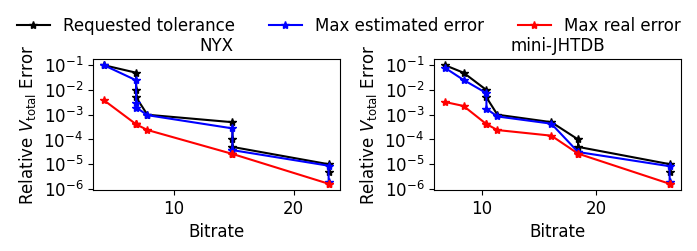}
    \vspace{-9mm}
    \caption{Request $V_{total}$ tolerance, max estimated $V_{total}$ errors, and max actual $V_{total}$ errors during progressive retrieval towards $V_{total}$ in the NYX and mini-JHTDB dataset}
    \label{fig:QoIControl}
    \vspace{-1em}
\end{figure}

\textit{Guaranteed QoI error control.} We validate the QoI error control by presenting and comparing three values: (1) requested tolerance $\tau$; (2) max estimated error computed by \ours; and (3) max real error of the provided data. 
As illustrated in Figure ~\ref{fig:QoIControl}, the max real error is always smaller than the max estimated error, which is close to but strictly smaller than the requested tolerance on both datasets. 
This shows that HP-MDR can faithfully enforce the QoI error control during progressive retrieval.


\subsubsection{Multi-GPU evaluation} 
\begin{figure}[t]
    \centering
    \includegraphics[width=0.85\columnwidth]{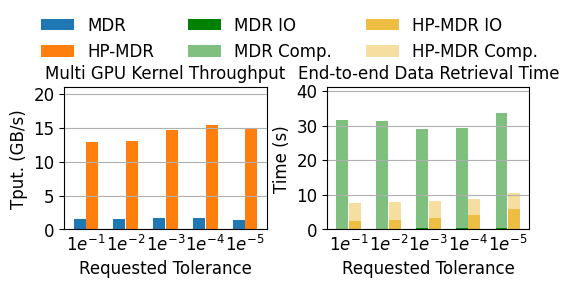}
    \vspace{-5mm}
    \caption{Multi GPU kernel throughput and end-to-end data retrieval time on the JHTDB dataset}
    \label{fig:QoI_TP_IO}
    \vspace{-2em}
\end{figure}

We further evaluate the throughput and end-to-end data retrieval performance of \ours on an entire Frontier node (8 GPUs), and compare it with multicore CPUs in the same node (64 cores) using the JHTDB dataset (48 GB). 
Under this setting, each CPU processes 0.75 GB of data, while each GPU handles 6 GB. 
We report both the overall kernel throughput (which only includes computational time) and the end-to-end data retrieval time (which measures the time from data reading to the completion of data reconstruction) in Fig.~\ref{fig:QoI_TP_IO}. 

According to the figure, \ours exhibits over $10.36\times$ speedup in kernel throughput, although the end-to-end performance gain is reduced to $4.18\times$. 
This is caused by two reasons: (1) I/O overhead is more significant in \ours because it creates many small files; and (2) there is some particular overhead in GPUs (e.g., memory allocation) for end-to-end evaluation. 
Nevertheless, we envision the I/O overhead could be mitigated with tailored implementation, which will further improve \ours's end-to-end performance. 

%% file: tex/conclusion.tex
\section{Conclusion}\label{sec:conclusion}
In this paper, we presented \ours, a high-performance and portable framework designed to accelerate data refactoring and progressive retrieval on heterogeneous systems with advanced GPUs. By thoroughly optimizing the bitplane encoding and lossless compression stages, we addressed key performance bottlenecks in current progressive methods. Our register block-based encoding and hybrid lossless compression techniques significantly improve throughput while maintaining data fidelity and portability. We further enhanced end-to-end efficiency through a pipeline optimization strategy that overlaps computation and memory operations. By integrating PMGARD and extending it with GPU-optimized QoI error control, \ours enables precise and efficient data retrieval tailored to the needs of scientific analytics. Extensive evaluations on real-world datasets across multiple GPU architectures demonstrated that \ours delivers substantial speedups over existing frameworks, achieving up to $6.6\times$ improvement in throughput and competitive retrieval efficiency. 
In the context of progressive retrieval under error constraints in derived Quantities of Interest, \ours leads to $10.4\times$ throughput for recomposing required data representations and $4.2\times$ performance for end-to-end  retrieval, when compared with state-of-the-art solutions. 

%% file: main.bbl

\begin{thebibliography}{40}


\ifx \showCODEN    \undefined \def \showCODEN     #1{\unskip}     \fi
\ifx \showDOI      \undefined \def \showDOI       #1{#1}\fi
\ifx \showISBNx    \undefined \def \showISBNx     #1{\unskip}     \fi
\ifx \showISBNxiii \undefined \def \showISBNxiii  #1{\unskip}     \fi
\ifx \showISSN     \undefined \def \showISSN      #1{\unskip}     \fi
\ifx \showLCCN     \undefined \def \showLCCN      #1{\unskip}     \fi
\ifx \shownote     \undefined \def \shownote      #1{#1}          \fi
\ifx \showarticletitle \undefined \def \showarticletitle #1{#1}   \fi
\ifx \showURL      \undefined \def \showURL       {\relax}        \fi
\providecommand\bibfield[2]{#2}
\providecommand\bibinfo[2]{#2}
\providecommand\natexlab[1]{#1}
\providecommand\showeprint[2][]{arXiv:#2}

\bibitem[aur({[n.\,d.]})]%
        {aurora}
 \bibinfo{year}{[n.\,d.]}\natexlab{}.
\newblock \bibinfo{title}{Aurora exscale system}.
\newblock \bibinfo{howpublished}{\url{https://www.alcf.anl.gov/support-center/aurora}}.
\newblock


\bibitem[elc({[n.\,d.]})]%
        {elcaptain}
 \bibinfo{year}{[n.\,d.]}\natexlab{}.
\newblock \bibinfo{title}{El Captain exscale system}.
\newblock \bibinfo{howpublished}{\url{https://asc.llnl.gov/exascale/el-capitan}}.
\newblock


\bibitem[fro({[n.\,d.]})]%
        {frontier}
 \bibinfo{year}{[n.\,d.]}\natexlab{}.
\newblock \bibinfo{title}{Frontier exscale supercomputer}.
\newblock \bibinfo{howpublished}{\url{https://www.olcf.ornl.gov/frontier}}.
\newblock


\bibitem[sum({[n.\,d.]})]%
        {summit}
 \bibinfo{year}{[n.\,d.]}\natexlab{}.
\newblock \bibinfo{title}{Summit exscale system}.
\newblock \bibinfo{howpublished}{\url{https://www.olcf.ornl.gov/summit}}.
\newblock


\bibitem[Ainsworth et~al\mbox{.}(2018)]%
        {ainsworth2018multilevel}
\bibfield{author}{\bibinfo{person}{Mark Ainsworth}, \bibinfo{person}{Ozan Tugluk}, \bibinfo{person}{Ben Whitney}, {and} \bibinfo{person}{Scott Klasky}.} \bibinfo{year}{2018}\natexlab{}.
\newblock \showarticletitle{Multilevel techniques for compression and reduction of scientific data—the univariate case}.
\newblock \bibinfo{journal}{\emph{Computing and Visualization in Science}} \bibinfo{volume}{19}, \bibinfo{number}{5} (\bibinfo{year}{2018}), \bibinfo{pages}{65--76}.
\newblock


\bibitem[Ainsworth et~al\mbox{.}(2019)]%
        {ainsworth2019multilevel}
\bibfield{author}{\bibinfo{person}{Mark Ainsworth}, \bibinfo{person}{Ozan Tugluk}, \bibinfo{person}{Ben Whitney}, {and} \bibinfo{person}{Scott Klasky}.} \bibinfo{year}{2019}\natexlab{}.
\newblock \showarticletitle{Multilevel techniques for compression and reduction of scientific data---the multivariate case}.
\newblock \bibinfo{journal}{\emph{SIAM Journal on Scientific Computing}} \bibinfo{volume}{41}, \bibinfo{number}{2} (\bibinfo{year}{2019}), \bibinfo{pages}{A1278--A1303}.
\newblock


\bibitem[Ainsworth et~al\mbox{.}(2020)]%
        {ainsworth2020multilevel}
\bibfield{author}{\bibinfo{person}{Mark Ainsworth}, \bibinfo{person}{Ozan Tugluk}, \bibinfo{person}{Ben Whitney}, {and} \bibinfo{person}{Scott Klasky}.} \bibinfo{year}{2020}\natexlab{}.
\newblock \showarticletitle{Multilevel techniques for compression and reduction of scientific data---The unstructured case}.
\newblock \bibinfo{journal}{\emph{SIAM Journal on Scientific Computing}} \bibinfo{volume}{42}, \bibinfo{number}{2} (\bibinfo{year}{2020}), \bibinfo{pages}{A1402--A1427}.
\newblock


\bibitem[Baker et~al\mbox{.}(2016)]%
        {baker2016evaluating}
\bibfield{author}{\bibinfo{person}{Allison~H Baker}, \bibinfo{person}{Dorit~M Hammerling}, \bibinfo{person}{Sheri~A Mickelson}, \bibinfo{person}{Haiying Xu}, \bibinfo{person}{Martin~B Stolpe}, \bibinfo{person}{Phillipe Naveau}, \bibinfo{person}{Ben Sanderson}, \bibinfo{person}{Imme Ebert-Uphoff}, \bibinfo{person}{Savini Samarasinghe}, \bibinfo{person}{Francesco De~Simone}, {et~al\mbox{.}}} \bibinfo{year}{2016}\natexlab{}.
\newblock \showarticletitle{Evaluating lossy data compression on climate simulation data within a large ensemble}.
\newblock \bibinfo{journal}{\emph{Geoscientific Model Development}} \bibinfo{volume}{9}, \bibinfo{number}{12} (\bibinfo{year}{2016}), \bibinfo{pages}{4381--4403}.
\newblock


\bibitem[Balevic(2009)]%
        {balevic2009fine}
\bibfield{author}{\bibinfo{person}{Ana Balevic}.} \bibinfo{year}{2009}\natexlab{}.
\newblock \bibinfo{title}{Fine-Grain Parallelization of Entropy Coding on GPGPUs}.
\newblock
\newblock


\bibitem[Ballester-Ripoll et~al\mbox{.}(2019)]%
        {ballester2019tthresh}
\bibfield{author}{\bibinfo{person}{Rafael Ballester-Ripoll}, \bibinfo{person}{Peter Lindstrom}, {and} \bibinfo{person}{Renato Pajarola}.} \bibinfo{year}{2019}\natexlab{}.
\newblock \showarticletitle{TTHRESH: Tensor compression for multidimensional visual data}.
\newblock \bibinfo{journal}{\emph{IEEE transactions on visualization and computer graphics}} \bibinfo{volume}{26}, \bibinfo{number}{9} (\bibinfo{year}{2019}), \bibinfo{pages}{2891--2903}.
\newblock


\bibitem[Cappello et~al\mbox{.}(2020)]%
        {cappello2020fulfilling}
\bibfield{author}{\bibinfo{person}{Franck Cappello}, \bibinfo{person}{Sheng Di}, {and} \bibinfo{person}{Ali~Murat Gok}.} \bibinfo{year}{2020}\natexlab{}.
\newblock \showarticletitle{Fulfilling the promises of lossy compression for scientific applications}. In \bibinfo{booktitle}{\emph{Driving Scientific and Engineering Discoveries Through the Convergence of HPC, Big Data and AI: 17th Smoky Mountains Computational Sciences and Engineering Conference, SMC 2020, Oak Ridge, TN, USA, August 26-28, 2020, Revised Selected Papers 17}}. Springer, \bibinfo{pages}{99--116}.
\newblock


\bibitem[Chen et~al\mbox{.}(2025)]%
        {chen2025hpdr}
\bibfield{author}{\bibinfo{person}{Jieyang Chen}, \bibinfo{person}{Qian Gong}, \bibinfo{person}{Yanliang Li}, \bibinfo{person}{Xin Liang}, \bibinfo{person}{Lipeng Wan}, \bibinfo{person}{Qing Liu}, \bibinfo{person}{Norbert Podhorszki}, {and} \bibinfo{person}{Scott Klasky}.} \bibinfo{year}{2025}\natexlab{}.
\newblock \showarticletitle{HPDR: High-Performance Portable Scientific Data Reduction Framework}.
\newblock \bibinfo{journal}{\emph{arXiv preprint arXiv:2503.06322}} (\bibinfo{year}{2025}).
\newblock


\bibitem[Chen et~al\mbox{.}(2021)]%
        {chen2021accelerating}
\bibfield{author}{\bibinfo{person}{Jieyang Chen}, \bibinfo{person}{Lipeng Wan}, \bibinfo{person}{Xin Liang}, \bibinfo{person}{Ben Whitney}, \bibinfo{person}{Qing Liu}, \bibinfo{person}{David Pugmire}, \bibinfo{person}{Nicholas Thompson}, \bibinfo{person}{Jong~Youl Choi}, \bibinfo{person}{Matthew Wolf}, \bibinfo{person}{Todd Munson}, {et~al\mbox{.}}} \bibinfo{year}{2021}\natexlab{}.
\newblock \showarticletitle{Accelerating multigrid-based hierarchical scientific data refactoring on gpus}. In \bibinfo{booktitle}{\emph{2021 IEEE International Parallel and Distributed Processing Symposium (IPDPS)}}. IEEE, \bibinfo{pages}{859--868}.
\newblock


\bibitem[Collet({[n.\,d.]})]%
        {zstd}
\bibfield{author}{\bibinfo{person}{Yann Collet}.} \bibinfo{year}{[n.\,d.]}\natexlab{}.
\newblock \bibinfo{title}{Zstandard - Real-time data compression algorithm}.
\newblock \bibinfo{howpublished}{\url{http://facebook.github.io/zstd/}}.
\newblock
\newblock
\shownote{Online}.


\bibitem[Deutsch(1996)]%
        {gzip}
\bibfield{author}{\bibinfo{person}{Peter Deutsch}.} \bibinfo{year}{1996}\natexlab{}.
\newblock \showarticletitle{GZIP file format specification version 4.3}.
\newblock  (\bibinfo{year}{1996}).
\newblock


\bibitem[Huang et~al\mbox{.}(2024)]%
        {huang2024cuszp2}
\bibfield{author}{\bibinfo{person}{Yafan Huang}, \bibinfo{person}{Sheng Di}, \bibinfo{person}{Guanpeng Li}, {and} \bibinfo{person}{Franck Cappello}.} \bibinfo{year}{2024}\natexlab{}.
\newblock \showarticletitle{cuSZp2: A GPU Lossy Compressor with Extreme Throughput and Optimized Compression Ratio}. In \bibinfo{booktitle}{\emph{Proceedings of the International Conference for High Performance Computing, Networking, Storage and Analysis}}. \bibinfo{pages}{1--18}.
\newblock


\bibitem[Huffman(1952)]%
        {huffman1952method}
\bibfield{author}{\bibinfo{person}{David~A Huffman}.} \bibinfo{year}{1952}\natexlab{}.
\newblock \showarticletitle{A method for the construction of minimum-redundancy codes}.
\newblock \bibinfo{journal}{\emph{Proceedings of the IRE}} \bibinfo{volume}{40}, \bibinfo{number}{9} (\bibinfo{year}{1952}), \bibinfo{pages}{1098--1101}.
\newblock


\bibitem[Ibarria et~al\mbox{.}(2003)]%
        {ibarria2003out}
\bibfield{author}{\bibinfo{person}{Lawrence Ibarria}, \bibinfo{person}{Peter Lindstrom}, \bibinfo{person}{Jarek Rossignac}, {and} \bibinfo{person}{Andrzej Szymczak}.} \bibinfo{year}{2003}\natexlab{}.
\newblock \showarticletitle{Out-of-core compression and decompression of large n-dimensional scalar fields}. In \bibinfo{booktitle}{\emph{Computer Graphics Forum}}, Vol.~\bibinfo{volume}{22}. Wiley Online Library, \bibinfo{pages}{343--348}.
\newblock


\bibitem[Lakshminarasimhan et~al\mbox{.}(2013)]%
        {lakshminarasimhan2013isabela}
\bibfield{author}{\bibinfo{person}{Sriram Lakshminarasimhan}, \bibinfo{person}{Neil Shah}, \bibinfo{person}{Stephane Ethier}, \bibinfo{person}{Seung-Hoe Ku}, \bibinfo{person}{Choong-Seock Chang}, \bibinfo{person}{Scott Klasky}, \bibinfo{person}{Rob Latham}, \bibinfo{person}{Rob Ross}, {and} \bibinfo{person}{Nagiza~F Samatova}.} \bibinfo{year}{2013}\natexlab{}.
\newblock \showarticletitle{ISABELA for effective in situ compression of scientific data}.
\newblock \bibinfo{journal}{\emph{Concurrency and Computation: Practice and Experience}} \bibinfo{volume}{25}, \bibinfo{number}{4} (\bibinfo{year}{2013}), \bibinfo{pages}{524--540}.
\newblock


\bibitem[Li et~al\mbox{.}(2019a)]%
        {li2019bstc}
\bibfield{author}{\bibinfo{person}{Ang Li}, \bibinfo{person}{Tong Geng}, \bibinfo{person}{Tianqi Wang}, \bibinfo{person}{Martin Herbordt}, \bibinfo{person}{Shuaiwen~Leon Song}, {and} \bibinfo{person}{Kevin Barker}.} \bibinfo{year}{2019}\natexlab{a}.
\newblock \showarticletitle{BSTC: A novel binarized-soft-tensor-core design for accelerating bit-based approximated neural nets}. In \bibinfo{booktitle}{\emph{Proceedings of the international conference for high performance computing, networking, storage and analysis}}. \bibinfo{pages}{1--30}.
\newblock


\bibitem[Li et~al\mbox{.}(2019b)]%
        {li2019vapor}
\bibfield{author}{\bibinfo{person}{Shaomeng Li}, \bibinfo{person}{Stanislaw Jaroszynski}, \bibinfo{person}{Scott Pearse}, \bibinfo{person}{Leigh Orf}, {and} \bibinfo{person}{John Clyne}.} \bibinfo{year}{2019}\natexlab{b}.
\newblock \showarticletitle{Vapor: A visualization package tailored to analyze simulation data in earth system science}.
\newblock \bibinfo{journal}{\emph{Atmosphere}} \bibinfo{volume}{10}, \bibinfo{number}{9} (\bibinfo{year}{2019}), \bibinfo{pages}{488}.
\newblock


\bibitem[Li et~al\mbox{.}(2023)]%
        {sperr}
\bibfield{author}{\bibinfo{person}{Shaomeng Li}, \bibinfo{person}{Peter Lindstrom}, {and} \bibinfo{person}{John Clyne}.} \bibinfo{year}{2023}\natexlab{}.
\newblock \showarticletitle{Lossy Scientific Data Compression With SPERR}. In \bibinfo{booktitle}{\emph{2023 IEEE International Parallel and Distributed Processing Symposium (IPDPS)}}. \bibinfo{pages}{1007--1017}.
\newblock
\urldef\tempurl%
\url{https://doi.org/10.1109/IPDPS54959.2023.00104}
\showDOI{\tempurl}


\bibitem[Liang et~al\mbox{.}(2018)]%
        {liang2018error}
\bibfield{author}{\bibinfo{person}{Xin Liang}, \bibinfo{person}{Sheng Di}, \bibinfo{person}{Dingwen Tao}, \bibinfo{person}{Sihuan Li}, \bibinfo{person}{Shaomeng Li}, \bibinfo{person}{Hanqi Guo}, \bibinfo{person}{Zizhong Chen}, {and} \bibinfo{person}{Franck Cappello}.} \bibinfo{year}{2018}\natexlab{}.
\newblock \showarticletitle{Error-controlled lossy compression optimized for high compression ratios of scientific datasets}. In \bibinfo{booktitle}{\emph{2018 IEEE International Conference on Big Data (Big Data)}}. IEEE, \bibinfo{pages}{438--447}.
\newblock


\bibitem[Liang et~al\mbox{.}(2021a)]%
        {liang2021error}
\bibfield{author}{\bibinfo{person}{Xin Liang}, \bibinfo{person}{Qian Gong}, \bibinfo{person}{Jieyang Chen}, \bibinfo{person}{Ben Whitney}, \bibinfo{person}{Lipeng Wan}, \bibinfo{person}{Qing Liu}, \bibinfo{person}{David Pugmire}, \bibinfo{person}{Rick Archibald}, \bibinfo{person}{Norbert Podhorszki}, {and} \bibinfo{person}{Scott Klasky}.} \bibinfo{year}{2021}\natexlab{a}.
\newblock \showarticletitle{Error-controlled, progressive, and adaptable retrieval of scientific data with multilevel decomposition}. In \bibinfo{booktitle}{\emph{Proceedings of the International Conference for High Performance Computing, Networking, Storage and Analysis}}. \bibinfo{pages}{1--13}.
\newblock


\bibitem[Liang et~al\mbox{.}(2021b)]%
        {liang2021mgard+}
\bibfield{author}{\bibinfo{person}{Xin Liang}, \bibinfo{person}{Ben Whitney}, \bibinfo{person}{Jieyang Chen}, \bibinfo{person}{Lipeng Wan}, \bibinfo{person}{Qing Liu}, \bibinfo{person}{Dingwen Tao}, \bibinfo{person}{James Kress}, \bibinfo{person}{David Pugmire}, \bibinfo{person}{Matthew Wolf}, \bibinfo{person}{Norbert Podhorszki}, {et~al\mbox{.}}} \bibinfo{year}{2021}\natexlab{b}.
\newblock \showarticletitle{Mgard+: Optimizing multilevel methods for error-bounded scientific data reduction}.
\newblock \bibinfo{journal}{\emph{IEEE Trans. Comput.}} \bibinfo{volume}{71}, \bibinfo{number}{7} (\bibinfo{year}{2021}), \bibinfo{pages}{1522--1536}.
\newblock


\bibitem[Liang et~al\mbox{.}(2022)]%
        {liang2022sz3}
\bibfield{author}{\bibinfo{person}{Xin Liang}, \bibinfo{person}{Kai Zhao}, \bibinfo{person}{Sheng Di}, \bibinfo{person}{Sihuan Li}, \bibinfo{person}{Robert Underwood}, \bibinfo{person}{Ali~M Gok}, \bibinfo{person}{Jiannan Tian}, \bibinfo{person}{Junjing Deng}, \bibinfo{person}{Jon~C Calhoun}, \bibinfo{person}{Dingwen Tao}, {et~al\mbox{.}}} \bibinfo{year}{2022}\natexlab{}.
\newblock \showarticletitle{Sz3: A modular framework for composing prediction-based error-bounded lossy compressors}.
\newblock \bibinfo{journal}{\emph{IEEE Transactions on Big Data}} \bibinfo{volume}{9}, \bibinfo{number}{2} (\bibinfo{year}{2022}), \bibinfo{pages}{485--498}.
\newblock


\bibitem[Lindstrom(2014)]%
        {lindstrom2014fixed}
\bibfield{author}{\bibinfo{person}{Peter Lindstrom}.} \bibinfo{year}{2014}\natexlab{}.
\newblock \showarticletitle{Fixed-rate compressed floating-point arrays}.
\newblock \bibinfo{journal}{\emph{IEEE transactions on visualization and computer graphics}} \bibinfo{volume}{20}, \bibinfo{number}{12} (\bibinfo{year}{2014}), \bibinfo{pages}{2674--2683}.
\newblock


\bibitem[Lindstrom et~al\mbox{.}(2025)]%
        {lindstrom2025zfp}
\bibfield{author}{\bibinfo{person}{Peter Lindstrom}, \bibinfo{person}{Jeffrey Hittinger}, \bibinfo{person}{James Diffenderfer}, \bibinfo{person}{Alyson Fox}, \bibinfo{person}{Daniel Osei-Kuffuor}, {and} \bibinfo{person}{Jeffrey Banks}.} \bibinfo{year}{2025}\natexlab{}.
\newblock \showarticletitle{ZFP: A compressed array representation for numerical computations}.
\newblock \bibinfo{journal}{\emph{The International Journal of High Performance Computing Applications}} \bibinfo{volume}{39}, \bibinfo{number}{1} (\bibinfo{year}{2025}), \bibinfo{pages}{104--122}.
\newblock


\bibitem[Lindstrom and Isenburg(2006)]%
        {lindstrom2006fast}
\bibfield{author}{\bibinfo{person}{Peter Lindstrom} {and} \bibinfo{person}{Martin Isenburg}.} \bibinfo{year}{2006}\natexlab{}.
\newblock \showarticletitle{Fast and efficient compression of floating-point data}.
\newblock \bibinfo{journal}{\emph{IEEE transactions on visualization and computer graphics}} \bibinfo{volume}{12}, \bibinfo{number}{5} (\bibinfo{year}{2006}), \bibinfo{pages}{1245--1250}.
\newblock


\bibitem[Liu et~al\mbox{.}(2021)]%
        {liu2021exploring}
\bibfield{author}{\bibinfo{person}{Jinyang Liu}, \bibinfo{person}{Sheng Di}, \bibinfo{person}{Kai Zhao}, \bibinfo{person}{Sian Jin}, \bibinfo{person}{Dingwen Tao}, \bibinfo{person}{Xin Liang}, \bibinfo{person}{Zizhong Chen}, {and} \bibinfo{person}{Franck Cappello}.} \bibinfo{year}{2021}\natexlab{}.
\newblock \showarticletitle{Exploring Autoencoder-Based Error-Bounded Compression for Scientific Data}.
\newblock \bibinfo{journal}{\emph{arXiv preprint arXiv:2105.11730}} (\bibinfo{year}{2021}).
\newblock


\bibitem[Magri and Lindstrom(2023)]%
        {magri2023general}
\bibfield{author}{\bibinfo{person}{Victor~AP Magri} {and} \bibinfo{person}{Peter Lindstrom}.} \bibinfo{year}{2023}\natexlab{}.
\newblock \showarticletitle{A general framework for progressive data compression and retrieval}.
\newblock \bibinfo{journal}{\emph{IEEE Transactions on Visualization and Computer Graphics}} \bibinfo{volume}{30}, \bibinfo{number}{1} (\bibinfo{year}{2023}), \bibinfo{pages}{1358--1368}.
\newblock


\bibitem[Pulido et~al\mbox{.}(2019)]%
        {pulido2019data}
\bibfield{author}{\bibinfo{person}{Jesus Pulido}, \bibinfo{person}{Zarija Lukic}, \bibinfo{person}{Paul Thorman}, \bibinfo{person}{Caixia Zheng}, \bibinfo{person}{James Ahrens}, {and} \bibinfo{person}{Bernd Hamann}.} \bibinfo{year}{2019}\natexlab{}.
\newblock \showarticletitle{Data reduction using lossy compression for cosmology and astrophysics workflows}. In \bibinfo{booktitle}{\emph{Journal of Physics: Conference Series}}, Vol.~\bibinfo{volume}{1290}. IOP Publishing, \bibinfo{pages}{012008}.
\newblock


\bibitem[Rabbani(2002)]%
        {rabbani2002jpeg2000}
\bibfield{author}{\bibinfo{person}{Majid Rabbani}.} \bibinfo{year}{2002}\natexlab{}.
\newblock \showarticletitle{JPEG2000: Image compression fundamentals, standards and practice}.
\newblock \bibinfo{journal}{\emph{Journal of Electronic Imaging}} \bibinfo{volume}{11}, \bibinfo{number}{2} (\bibinfo{year}{2002}), \bibinfo{pages}{286}.
\newblock


\bibitem[Tao et~al\mbox{.}(2017)]%
        {sz17}
\bibfield{author}{\bibinfo{person}{Dingwen Tao}, \bibinfo{person}{Sheng Di}, \bibinfo{person}{Zizhong Chen}, {and} \bibinfo{person}{Franck Cappello}.} \bibinfo{year}{2017}\natexlab{}.
\newblock \showarticletitle{Significantly improving lossy compression for scientific data sets based on multidimensional prediction and error-controlled quantization}. In \bibinfo{booktitle}{\emph{2017 IEEE International Parallel and Distributed Processing Symposium}}. IEEE, \bibinfo{pages}{1129--1139}.
\newblock


\bibitem[Tian et~al\mbox{.}(2020)]%
        {tian2020cusz}
\bibfield{author}{\bibinfo{person}{Jiannan Tian}, \bibinfo{person}{Sheng Di}, \bibinfo{person}{Kai Zhao}, \bibinfo{person}{Cody Rivera}, \bibinfo{person}{Megan~Hickman Fulp}, \bibinfo{person}{Robert Underwood}, \bibinfo{person}{Sian Jin}, \bibinfo{person}{Xin Liang}, \bibinfo{person}{Jon Calhoun}, \bibinfo{person}{Dingwen Tao}, {and} \bibinfo{person}{Franck Cappello}.} \bibinfo{year}{2020}\natexlab{}.
\newblock \showarticletitle{Cusz: An efficient gpu-based error-bounded lossy compression framework for scientific data}. In \bibinfo{booktitle}{\emph{Proceedings of the ACM International Conference on Parallel Architectures and Compilation Techniques}}. \bibinfo{pages}{3--15}.
\newblock


\bibitem[Tian et~al\mbox{.}(2021)]%
        {tian2021revisiting}
\bibfield{author}{\bibinfo{person}{Jiannan Tian}, \bibinfo{person}{Cody Rivera}, \bibinfo{person}{Sheng Di}, \bibinfo{person}{Jieyang Chen}, \bibinfo{person}{Xin Liang}, \bibinfo{person}{Dingwen Tao}, {and} \bibinfo{person}{Franck Cappello}.} \bibinfo{year}{2021}\natexlab{}.
\newblock \showarticletitle{Revisiting huffman coding: Toward extreme performance on modern gpu architectures}. In \bibinfo{booktitle}{\emph{2021 IEEE International Parallel and Distributed Processing Symposium (IPDPS)}}. IEEE, \bibinfo{pages}{881--891}.
\newblock


\bibitem[Underwood et~al\mbox{.}(2024)]%
        {underwood2024understanding}
\bibfield{author}{\bibinfo{person}{Robert Underwood}, \bibinfo{person}{Jon~C Calhoun}, \bibinfo{person}{Sheng Di}, {and} \bibinfo{person}{Franck Cappello}.} \bibinfo{year}{2024}\natexlab{}.
\newblock \showarticletitle{Understanding The Effectiveness of Lossy Compression in Machine Learning Training Sets}.
\newblock \bibinfo{journal}{\emph{arXiv preprint arXiv:2403.15953}} (\bibinfo{year}{2024}).
\newblock


\bibitem[Wallace(1992)]%
        {wallace1992jpeg}
\bibfield{author}{\bibinfo{person}{Gregory~K Wallace}.} \bibinfo{year}{1992}\natexlab{}.
\newblock \showarticletitle{The JPEG still picture compression standard}.
\newblock \bibinfo{journal}{\emph{IEEE transactions on consumer electronics}} \bibinfo{volume}{38}, \bibinfo{number}{1} (\bibinfo{year}{1992}), \bibinfo{pages}{xviii--xxxiv}.
\newblock


\bibitem[Wu et~al\mbox{.}(2024)]%
        {wu2024error}
\bibfield{author}{\bibinfo{person}{Xuan Wu}, \bibinfo{person}{Qian Gong}, \bibinfo{person}{Jieyang Chen}, \bibinfo{person}{Qing Liu}, \bibinfo{person}{Norbert Podhorszki}, \bibinfo{person}{Xin Liang}, {and} \bibinfo{person}{Scott Klasky}.} \bibinfo{year}{2024}\natexlab{}.
\newblock \showarticletitle{Error-controlled Progressive Retrieval of Scientific Data under Derivable Quantities of Interest}. In \bibinfo{booktitle}{\emph{SC24: International Conference for High Performance Computing, Networking, Storage and Analysis}}. IEEE, \bibinfo{pages}{1--16}.
\newblock


\bibitem[Zhao et~al\mbox{.}(2021)]%
        {zhao2021optimizing}
\bibfield{author}{\bibinfo{person}{Kai Zhao}, \bibinfo{person}{Sheng Di}, \bibinfo{person}{Maxim Dmitriev}, \bibinfo{person}{Thierry-Laurent~D Tonellot}, \bibinfo{person}{Zizhong Chen}, {and} \bibinfo{person}{Franck Cappello}.} \bibinfo{year}{2021}\natexlab{}.
\newblock \showarticletitle{Optimizing error-bounded lossy compression for scientific data by dynamic spline interpolation}. In \bibinfo{booktitle}{\emph{2021 IEEE 37th International Conference on Data Engineering (ICDE)}}. IEEE, \bibinfo{pages}{1643--1654}.
\newblock


\end{thebibliography}
